\documentclass[prd,aps,preprint,amsmath,nofootinbib,amssymb,eqsecnum,showkeys,tightenlines]{revtex4-1}
\pdfoutput=1
\usepackage{tabularx}
\usepackage{slashed}
\usepackage{epsfig,latexsym,cancel,amssymb,amsmath,verbatim,mathrsfs}
\usepackage{color}
\usepackage{graphicx}
\usepackage{bm}
\usepackage{tabularx}
\usepackage{amsmath}
\usepackage{amssymb}
\usepackage{amsfonts}
\usepackage{cases}
\usepackage{cancel}
\usepackage{hyperref}
\usepackage{ulem}
\usepackage{here}
\usepackage{color}
\usepackage{graphicx}
\usepackage{diagbox}
\usepackage{multirow}
\usepackage{slashed}
\usepackage{simpler-wick}
\usepackage{subfigure}
\usepackage{svg}
\usepackage[marginal]{footmisc}
\usepackage{feynmf}

\def\vec{\mathbf}

\def\wt{\widetilde}

\newcommand{\be}{\begin{equation}}
\newcommand{\ee}{\end{equation}}
\newcommand{\bea}{\begin{eqnarray}}
\newcommand{\eea}{\end{eqnarray}}
\newcommand{\ba}{\begin{array}}
\newcommand{\ea}{\end{array}}

\def\wt{\widetilde}

\long\def\symbolfootnote[#1]#2{\begingroup%
\def\thefootnote{\fnsymbol{footnote}}\footnote[#1]{#2}\endgroup}

\DeclareMathOperator{\Tr}{Tr} \DeclareMathOperator{\diag}{diag}

\newcommand{\beq}{\begin{equation}}
\newcommand{\eeq}{\end{equation}}

\begin{document}

\title{Confinement and Chiral Phase Transitions: The Role of Polyakov Loop Kinetics Terms}

\author{Banghui Hua}
\email{bhhua@hust.edu.cn}
\affiliation{School of Physics, Huazhong University of Science and Technology, Wuhan 430074, China}

\author{Zhaofeng Kang}
\email{zhaofengkang@gmail.com}
\affiliation{School of physics, Huazhong University of Science and Technology, Wuhan 430074, China}

\author{Jiang Zhu}
\email{jackpotzhujiang@gmail.com}
\affiliation{Tsung-Dao Lee Institute and  School of Physics and Astronomy, Shanghai Jiao Tong University,
800 Lisuo Road, Shanghai, 200240 China}
\affiliation{Shanghai Key Laboratory for Particle Physics and Cosmology, 
Key Laboratory for Particle Astrophysics and Cosmology (MOE), 
Shanghai Jiao Tong University, Shanghai 200240, China}

\date{\today}

\begin{abstract}
We studied a crucial but often oversimplified ingredient in predicting gravitational-wave signals from QCD-type phase transitions: the kinetic term of the Polyakov loop. For the first time, we derive this term from first principles in finite-temperature pure SU(3) Yang-Mills theory, incorporating a field-dependent renormalization factor—a calculation we also extend to theories with more colors. Employing this derived kinetic term alongside three commonly-used effective potentials (the Haar-measure, polynomial, and quasi-particle models), we demonstrate that it substantially modifies the predicted GW energy spectrum from confinement transitions by 1-2 orders of magnitude. Based on this, we provide the first complete analysis of the chiral transition within the Polyakov–Nambu–Jona-Lasinio (PNJL) framework, described by the quark condensate. Our results reveal a clear dichotomy: while the Polyakov-loop kinetic term critically shapes GWs from confinement transitions, it has a negligible impact on the dynamics of the chiral transition, which is dominated by fermion condensation effects.

\end{abstract}


\maketitle

\section{Introduction}\label{sec:intro}

Cosmological first-order phase transitions are a prime target for upcoming gravitational-wave observatories, as they could generate a stochastic background encoding vital information about fundamental particle physics~\cite{Apreda:2001us, Leitao:2012tx, Gould:2019qek, Kang:2020jeg, Friedrich:2022cak, NANOGrav:2023hvm}. Among them, the QCD-type phase transition—encompassing both confinement and chiral symmetry breaking—serves as a critical benchmark of the strong dynamics in the standard model. Confinement, governed by the pure Yang-Mills sector, is a defining feature of the Standard Model and may also occur in dark sectors of Beyond the Standard Model physics~\cite{Beylin:2020bsz, Halverson:2020xpg, Huang:2020crf, Kang:2021epo, Morgante:2022zvc, Fujikura:2023lkn, Li:2023xto, He:2023ado, Sagunski:2023ynd, Kang:2025nhe}. A precise theoretical description of this transition is therefore essential for connecting particle physics to cosmological observables.

Similar to cosmological first-order transitions, a first-order QCD phase transition proceeds via the nucleation and expansion of true vacuum bubbles~\cite{Hindmarsh:2020hop, Arunasalam:2021zrs, Lofgren:2021ogg, Caldwell:2022qsj}. A crucial quantity governing this dynamics is the bubble nucleation rate per unit volume (equivalent to the false vacuum decay rate), as it directly determines the transition history and the resulting gravitational wave spectrum~\cite{Enqvist:1991xw, Kamionkowski:1993fg, Caprini:2009yp, Hindmarsh:2013xza, Jinno:2016vai, Jinno:2017fby, Jinno:2017ixd, Caprini:2019egz, Lewicki:2020azd}. Computing this rate requires knowledge of the effective potential and the kinetic term for the relevant order parameter. For QCD-type transitions, however, the order parameters are composite fields of the strong dynamics, making their non-perturbative effective potential and kinetic term notoriously difficult to derive.

Since the finite-temperature effective potential is directly related to thermodynamic quantities, lattice thermodynamic data can directly guide the construction of such effective potentials. Consequently, many studies have conducted detailed discussions on how to construct the non-perturbative effective potential for QCD phase transitions. The discussions mainly focus on three approaches: 1) constructing effective models for the order parameter based on symmetry considerations~\cite{Pisarski:2000eq, Pisarski:2001pe, Dumitru:2003hp, Lucini:2005vg, Ratti:2006gh, Roessner:2006xn, Fukushima:2013rx, Fukushima:2017csk, Huang:2020crf, Halverson:2020xpg, Kang:2021epo}, which is also the main focus in this paper; 2) computing the corresponding effective potential using functional renormalization group (FRG) evolution~\cite{Welzbacher:2014pea, Fister:2015eca, Eser:2015pka, Gao:2015kea, Cyrol:2017ewj, Cyrol:2017qkl, Fu:2019hdw, Gao:2020qsj, Lu:2023mkn, Gao:2024ggp, Guan:2025mce, Lu:2025cls}; and 3) using the Ads/QCD holographic mapping methods~\cite{Arefeva:2018hyo, Fang:2018axm, Liu:2023pbt}. All three approaches require fitting model parameters using lattice data, ultimately matching the temperature-dependent evolution of lattice thermodynamic observables.

However, the situation is fundamentally opposite for the kinetic term. Due to its lack of a direct thermodynamic counterpart, there is no guiding input from lattice data~\footnote{The only relative lattice quantities is the two-point functions of the gauge filed or the Polyakov loops~\cite{Dittmann:2003qt}. }. Its computation can only rely on specific physical models and processes. For the confinement phase transition involving pure Yang-Mills fields, the order parameter is the trace of the Polyakov loop, a composite field constructed from temporal gauge fields. Consequently, constructing its kinetic term is highly non-trivial; Adrian Dumitrua and Robert D. Pisarski even found that it exhibits a temperature-dependent kinetic term at the tree level~\cite{Dumitru:2000in}, and applied in many papers~\cite{Meisinger:2001cq,Meisinger:2001fi,Scavenius:2002ru,Pisarski:2006hz} or even 1-loop level~\cite{Wirstam:2001ka}. Recently, Stephan Huber and his coworkers found lattice data suggest that the kinetic term of Polyakov loop derived from the scalar cases significantly~\cite {Huber:2025qbl}. Nevertheless, in discussions concerning bubble nucleation and associated gravitational wave signals in the early universe, the effect of the non-trivial kinetic term is often neglected, typically assuming a canonical form~\cite{Halverson:2020xpg, Kang:2021epo, Reichert:2022naa}. Therefore, the precise impact of the kinetic term on the physical observables predicted by such models remains unknown.
Regarding the chiral phase transition involving fermions, the directly relevant order parameter is the fermion chiral condensate. Within a given effective model, the kinetic term for this transition can, to some extent, be addressed perturbatively. However, due to the composite nature of the order parameter, its kinetic term exhibits non-trivial features. The work in~\cite{Helmboldt:2019pan} was the first to discuss this effect using the NJL model. They found that without radiative corrections, the kinetic term for the order parameter is zero, whereas including one-loop thermal corrections generates a non-trivial field-dependent kinetic term. Many subsequent studies have discussed possible chiral phase transitions in the early universe based on this framework~\cite{Reichert:2021cvs, Sagunski:2023ynd, Kang:2025nhe}, and even including the non-zero chemical potential which would significantly enhance the GWs~\cite{Kang:2025nhe}.
However, a crucial question remains to be clarified in such discussions. Since the QCD chiral phase transition involves simultaneous changes in multiple order parameters, namely the fermion chiral condensate and the trace of the Polyakov loop, what impact does the unknown kinetic term of the Polyakov loop have on the chiral phase transition and its observable signals? ~\footnote{In our previous works, we have briefly discussed these problems without the renormalization factor of the Polyakov loop~\cite{Hua:2025fap}.} 
This is one of the central questions this paper aims to answer. Therefore, starting from the confinement phase transition in pure Yang-Mills theory, this work will systematically construct the kinetic term for its order parameter and investigate how it affects bubble nucleation and gravitational wave production. Building on this foundation, we will then incorporate fermions to ultimately address this question.

This paper is organized as follows: In section.\ref{ACTION}, we would briefly introduce how to compute the nucleation rate of the phase transition from the effective action. We, for the first time, systematically derived the non-trivial tree-level kinetic term of the confinement phase transition order parameters in section.\ref{Conf}. To demonstrate the importance of those quantities, we provided an analysis of how the non-trivial kinetic terms affect the GW spectrum from the confinement phase transition for various effective models.
In the section.\ref{LAG}, we will add back the fermion contribution and reanalyze the effect of the non-trivial kinetic terms both for the Polyakov loop and the fermion condensation. Conclusions and discussions, as well as the appendix, are cast in the remaining two sections.

\section{The Bubble Nucleation rate and the Renormalization Factors}\label{ACTION}
Just like the discussion in the introduction, the bubble nucleation rate/vacuum decay rate is the central physical quantity if we want to discuss the observables from the early universe phase transitions. This rate can be defined as $\Gamma=-\frac{1}{P(t)}\frac{dP(t)}{dt}|_{t=t_n}$ where $t_n$ in the nucleation times and the $P( t)=|\langle\phi=\phi_{fv}|e^{-iHt}|\phi=\phi_{fv}\rangle|^2$ is the probability that universe at the false vacuum. Analogously to how a particle's decay rate is given by the imaginary part of its energy, the nucleation rate can be derived from the imaginary part of the generating functional computed via the path integral~\cite{Coleman:1977py,Callan:1977pt,Linde:1977mm,Linde:1980tt,Linde:1981zj, Weinberg:1992ds, Andreassen:2014eha,Ai:2019fri,Ai:2020sru,Hirvonen:2020jud,Lofgren:2021ogg,Hirvonen:2021zej,Gould:2021ccf,Hirvonen:2024rfg}. As a consequence, the expression of the nucleation rate per volume can be formulated as an $\mathcal{O}(1)$ coefficient of the functional determinant times an exponential term governed by the effective action of the tunneling field(the order parameters)
\begin{equation}
\begin{split}
    \frac{\Gamma}{V}&=\left(\frac{S_E[\phi_b]}{2\pi}\right)^{d/2}\left|\frac{\det'[\delta^2 S_E/\delta\phi^2]_{\phi=\phi_b}}{\det[\delta^2 S_E/\delta\phi^2]_{\phi=\phi_{fv}}}\right|e^{-S_E[\phi_b]-S_E[\phi_{fv}]}.
\end{split}
\end{equation}
In the above expressions, the $\phi_b$, known as the bounce solution, is defined as the solution to the EoMs from the effective action that obeys the bounce boundary conditions. Now, let us dive deeper into this solution, which is needed if we want to compute a numerical value of the nucleation rate. To obtain the bounce solution, let us first focus on the effective action itself, which, by definition, is the generating functional of the $ n$-point functions
\begin{equation}
\begin{split}
S_{eff}(\phi_c)=\sum_{n=0}^\infty\frac{1}{n!} \int d^4x_1...d^4 x_n\bar{\phi}(x_1)...\bar{\phi}(x_n)\Gamma^{(n)}(x_1,...,x_n).    
\end{split}
\end{equation}
One can perform the derivative expansion of the background field $\bar{\phi}$ up to the next leading order to obtain a more familiar form~\cite{Quiros:1999jp, Hua:2025fap}
\begin{equation}\label{eq:anticeffac}
    S_{eff}(\bar{\phi})=\int d^4 x [-V_{eff}(\bar{\phi})+\frac{Z(\bar{\phi})}{2}\partial_\mu\bar{\phi}\partial^\mu\bar{\phi}+...].
\end{equation}
The first and second terms can be recognized as the effective potential and the kinetic term of the background field $\bar\phi$. As one can observe that the kinetic terms always admit a coefficient $Z(\bar\phi)$, the renormalization factor of $\bar\phi$, from derivative expansion, and by definition, it could be obtained as
\begin{equation}
    Z(\bar\phi)=-\frac{\partial\wt\Gamma^{(2)}(p^2)}{\partial p^2}|_{p^2=0},
\end{equation}
where $\wt\Gamma^{(2)}(p^2)$ is the 2-point function at the momentum space. One can always separate $Z(\bar\phi)$ into the classical contribution $Z_T$ and the quantum correction $Z_Q$ as following
\begin{equation}
    Z(\bar\phi)=-\bigg[\frac{\partial\wt\Gamma_{T}^{(2)}(p^2)}{\partial p^2}|_{p^2=0}+\hbar\frac{\partial\wt\Gamma_{Q}^{(2)}(p^2)}{\partial p^2}|_{p^2=0}\bigg].
\end{equation}
For elementary fields, such as the Higgs for the electroweak phase transition, only the loop correction is non-trivial. However, for composite fields, such as the quark condensate on the chiral phase transition, both $Z_T$ and $Z_Q$ admit a non-zero contribution~\cite{Helmboldt:2019pan,Reichert:2021cvs, Sagunski:2023ynd, Kang:2025nhe}. Furthermore, even for the elementary fields, the non-trivial contribution from $Z_Q$ plays an important role if we want to make the nucleation rate gauge independent~\cite{Lofgren:2021ogg,Hirvonen:2021zej}. 

In summary, the renormalization factor $Z(\phi)$ plays an indispensable role in the calculation of phase transition dynamics. Its accurate computation is essential, both for maintaining gauge independence in theories with fundamental fields (e.g., the electroweak phase transition) and for correctly capturing the behavior of composite order parameters, such as the quark condensate in chiral phase transitions.

In the following chapter, we will apply the above general framework to the simplest system in QCD-like theories—the pure Yang-Mills theory—to concretely illustrate the physical impact of the order parameter renormalization factor. We focus on this system for two primary reasons: First, its phase transition is governed by confinement, for which the order parameter, the Polyakov loop $l$, is a quintessential composite field. Second, and more importantly, unlike the chiral case where non-triviality arises at the loop level, the confinement phase transition exhibits a non-trivial tree-level contribution to the renormalization factor, $Z_T(l)$. This provides an ideal and clearer framework to first demonstrate the significant effects of $Z(\phi)$ without the immediate complication of dominant quantum corrections.

\section{The effect of the PL kinetic terms inside the confinement phase transition}\label{Conf}
The confinement phase transition in the $SU(N_c)$ pure Yang-Mills theory admits an order parameter, which is associated with the $Z_{N_c}$ center symmetry breaking, the trace of the Polyakov loop
\begin{equation}\label{define:Pol}
    l={\Tr_c}L=\frac{\Tr_c}{N}{\cal P}\exp\left(-ig\int_0^\beta A^a_4T^ad\tau\right).
\end{equation}
where $\mathcal{P}$ indicated the path order and $\beta$ is the inversed temperature. This quantity is related to the exponential of the minus free energy $\exp[-F(T)]$ which mean that $l\rightarrow0$ in the confinement phase($F(T)\rightarrow\infty$), but $l\rightarrow1$ in the fully deconfinement phase($\lim_{T\rightarrow\infty}F(T)\rightarrow0$). As we mentioned in the previous section, to compute the true vacuum nucleation rate of the order parameter, both the kinetic terms and the effective potential are essential. Those two terms would determine the bounce solution and affect the 3-dimensional Euclidean action. In principle, one needs to derive both of them from the first principle lagrangian, but none of these requirements can be satisfied due to the strongly coupled problem around the confinement phase transition. As a consequence, one can only start to construct the effective theory of confinement based on the order parameters. The effective potential of the order parameters is relatively easy to construct compared with the kinetic energy term. This is because the effective potential at the true vacuum is related to the thermal pressure, which can be computed by the lattice QCD from first principles. In other words, one can construct the effective potential based on the lattice data. As an example, one can construct the effective potential from the Landau-Ginzburg theory and fit the coefficient from the lattice data~\cite{Ratti:2006gh, Roessner:2006xn}. 

On the other hand, the kinetic term is trickier. There are two main problems that hinder the construction of the kinetic term for the Polyakov loops: 
\begin{itemize}
    \item {\bf Composite Nature}: the Polyakov loop itself is a composite object which constructed by the elementary field $A_4$.
    \item {\bf Quantum Correction}: just like the effective potential, the kinetic terms itself also governed by the strong dynamics, but the directly related lattice quantities do not exist.
\end{itemize} 
Based on these two points, it is quite difficult to construct the kinetic terms of the Polyakov Loops. As a compromise method, the previous study normally treats $l$ as an elementary scalar field, which admits a kinetic term as
$\vec{\nabla} l^\dagger\cdot\vec{\nabla}l$. But we mentioned in  Eq.\eqref{eq:anticeffac}, the general form of the kinetic terms would admit a renormalization factor $Z(l)$, which would be affected by both the Composite Nature of the Polyakov Loop and the Quantum correction from the strong dynamics. As we introduced in Sec.~\ref{sec:intro}, Adrian Dumitrua and Robert D. Pisarski found the Polyakov loop should have a temperature-dependent kinetic term at the tree level form as~\cite{Dumitru:2000in} $Z_l\propto T^2\frac{N}{g^2}$. However, just like what they briefly explained in that paper, those kinetic terms are obtained by a "reasonable guess" based on the kinetic terms of spatial Wilson line and the Lorentz invariant. As a result, the explicit form of the $Z_l$ is remain unknown. In the following discussion, we would like to show how to construct $Z_l(l)$, which includes the Composite Nature of the Polyakov Loop at least at the tree level, and how important the non-trivial $Z(l)$ is.


\subsection{Renormalization factor for PYM}
Despite the fact that previous studies of the confinement phase transition have often considered the quantum corrections encoded by the lattice data, we would like to start the discussions on the kinetic terms through the simplest quantities that govern the pure Yang-Mills theory:  tree-level Lagrangian. One can firstly decompose the classical Lagrangian into two parts
\begin{equation}
    \mathcal{L}=-\frac{1}{2}\mathrm{Tr}\left(F_{\mu\nu}F^{\mu\nu}\right)
    =K-V.
\end{equation}
The kinetic term arises from the terms that control the propagation of the field. It is composed of the quadratic terms in the derivatives of the gauge field $A_\mu$
\begin{equation}
\begin{split}
    K&=-\frac{1}{2}\mathrm{Tr}(2\partial_\mu A_0 \partial^\mu A^0 -2\partial_0 A_\mu \partial^0 A^\mu).
\end{split}
\end{equation}
The other part of the Lagrangian that represents the interaction, which 
contains the cubic and quartic terms of $A$, constitutes the potential term, denoted as $V$,
\begin{equation}
\begin{split}
    V=g^2f^{abc}f^{ab'c'}A_\mu^b A_\nu^c A^{\mu b'}A^{\mu c'}+2gf^{abc}A_\mu^b A_\nu^c \partial^\mu A^\nu -2gf^{abc}A_\mu^b A_\nu^c\partial_\nu A_\mu^a.
\end{split}
\end{equation}
Since the order parameter of the confinement phase transition is the Polyakov loop, which involves only the temporal component of the field, $A_\mu$ in the Lagrangian should also contain only a temporal component, i.e., $A_\mu=A_0 \delta_{\mu 0}$. Under this gauge, one can prove that the tree-level potential must be zero~\cite{Kang:2024xqk}. Moreover, with the help of an effective theory, such as the quasi-particle model, one can prove that the effective potential of confinement is entirely generated by the finite temperature quantum corrections~\cite{Sasaki:2012bi, Reinosa:2014ooa,vanEgmond:2021jyx,Kang:2022jbg}. The gauge choice also simplifies the kinetic term to
\begin{equation}\label{termkaa}
\begin{split}
    K=\mathrm{Tr}\left(\bm{\nabla} A_0 \cdot\bm{\nabla} A_0 \right)
    =-\mathrm{Tr}\left(\bm{\nabla} A_4 \cdot\bm{\nabla} A_4 \right),
\end{split}
\end{equation}
where we have used the Euclidean time formalism and identified $A_0$ with $A_4$. Since we aim to identify the kinetic term of the order parameter, we need to express the above kinetic term in terms of the Polyakov loop. On the other hand, the Polyakov loop is defined as
\begin{equation}\label{lsu3}
    L=\frac{1}{3}e^{-i\frac{A_4}{T}}.
\end{equation}
As we know that the kinetic term of the order parameter is composed of the quadratic terms in the derivatives of it, we must link $\bm{\nabla}L$ to $\bm{\nabla}A_4$. For comparison with the scalar field case, whose kinetic term is written as $\bm{\nabla} \phi^\dagger \cdot \bm{\nabla} \phi$, we intend to express that for the Yang-Mills gauge field in terms of $L$ and its Hermitian conjugate. For color number $N=3$, we have
\begin{align}
    L^{\dagger}L=\frac{1}{3}e^{i\frac{A^\dagger_4}{T}}\frac{1}{3}e^{-i\frac{A_4}{T}}=\frac{1}{9}.
\end{align}
To differentiate both sides, we can obtain $\bm{\nabla}(L^\dagger L)=0$. Thus, we can obtain the relation between $\bm{\nabla}L^\dagger$ and $\bm{\nabla}L$,
\begin{equation}\label{eq:lnl}
    (\bm{\nabla}L^\dagger)L+L^\dagger\bm{\nabla}L=0 ~~\Rightarrow ~~ L^\dagger\bm{\nabla}L=-(\bm{\nabla}L^\dagger)L.
\end{equation}
On the other hand, the spatial derivative of $L$ is given by $\bm{\nabla}L=-\frac{i}{T}L\bm{\nabla}A_4$. This allows us to express the derivative of $A_4$ in terms of $L$ as
\begin{equation}\label{eq:na4}
    \bm{\nabla}A_4=9iTL^\dagger\bm{\nabla}L.
\end{equation}
Combining Eq.\eqref{eq:na4} and Eq.\eqref{eq:lnl}, the gradient squared of $A_4$ can be expressed in terms of $L$ and $L^\dagger$:
\begin{equation}
\begin{split}
    \bm{\nabla}A_4\cdot\bm{\nabla}A_4&=-(iT)^2\left[9(\bm{\nabla}L^\dagger )L\right]\cdot\left[9L^\dagger\bm{\nabla}L\right]
    \\
    &=9T^2\bm{\nabla}L^\dagger\cdot\bm{\nabla}L,
\end{split}
\end{equation}
and the kinetic term of the background field as in Eq.\eqref{termkaa} is given by
\begin{equation}
\begin{split}
    K=-9T^2\mathrm{Tr}\left(\bm{\nabla}L^\dagger\cdot\bm{\nabla}L\right).
\end{split}
\end{equation}
As a result, the Lagrangian expressed by the order parameter is
\begin{equation}
    \mathcal{L}=-9T^2\mathrm{Tr}\left(\bm{\nabla}L^\dagger\cdot\bm{\nabla}L\right)-V_{PLM},
\end{equation}
where $V_{PLM}$ is the effective potential of the Polyakov loop, which would be introduced later. Since we aim to study the confinement phase transition, we need to compute the corresponding Euclidean action, which can be written as
\begin{equation}
    S_E=\beta\int d^3 \vec{x}\left[9T^2\mathrm{Tr}\left(\bm{\nabla}L^\dagger\cdot\bm{\nabla}L\right)+V_{PLM}\right].
\end{equation}
We prefer to express the kinetic term as a scalar rather than a matrix, so we compute the trace in it. First, write the matrix form of $L$. For $SU(3)$ theory, one can write the Polyakov loop by the eigen phase as ${L}=\diag[e^{ i(\frac{A_4^3}{2T}+\frac{A_4^8}{2\sqrt{3}T})}, e^{i(\frac{-A_4^3}{2T}+\frac{A_4^8}{2\sqrt{3}T})},e^{i(-\frac{A_4^8}{\sqrt{3}T})}]$. This expression gives $\sum_i q_i =0$, when we write $L$ as
\begin{equation}\label{matex}
    L=\frac{1}{3}
    \begin{pmatrix}
    e^{iq_1} &          &          \\
             & e^{iq_2} &          \\
             &          & e^{iq_3}
    \end{pmatrix}.
\end{equation}
Since we require its trace to be real, in the Polyakov loop gauge we have $q_1=-q_3=q$, $q_2=0$, which is called equal-eigenvalue condition. The proof of this condition can be found in Append.\ref{appA}. Thus, we can express the derivative of $L$ and its Hermitian conjugate by one parameter $q$, and the kinetic term can be found by multiply $\bm{\nabla}L$ and its Hermitian conjugate as follow
\begin{equation}
    \bm{\nabla}L=\frac{1}{3}
    \begin{pmatrix}
    ie^{iq}\bm{\nabla}q &   &                     \\
                        & 0 &                     \\
                        &   & -ie^{-iq}\bm{\nabla}q
    \end{pmatrix}, \bm{\nabla}L^\dagger\cdot\bm{\nabla}L=\frac{1}{9}
    \begin{pmatrix}
    \bm{\nabla}q\cdot\bm{\nabla}q &   &                               \\
                                  & 0 &                               \\
                                  &   & \bm{\nabla}q\cdot\bm{\nabla}q
    \end{pmatrix}.
\end{equation}
Therefore, the Euclidean action becomes $S_E=\beta\int d^3\vec{x}\left(2T^2\bm{\nabla}q\cdot\bm{\nabla}q+V_{PLM}\right)$.
We finally need to express the Euclidean action by the trace of the Polyakov loop, $l$, which is defined in Eq.\eqref{define:Pol}. So, we write $l$ in term of $q$ as 
\begin{equation}\label{eq:eeconif}
\begin{split}
    l=\frac{1}{3}\mathrm{Tr}
    \begin{pmatrix}
        e^{iq}&   &       \\
              & 1 &       \\
              &   &e^{-iq}
    \end{pmatrix}
    =\frac{1}{3}(1+2\cos q),
\end{split}
\end{equation}
hence, $q$ can be expressed in terms of 
$l$ using the inverse function, $    q=\arccos \frac{3l-1}{2},\  \  
    \frac{dq}{dl}=\frac{-3}{2\sqrt{1-\left(\frac{3l-1}{2}\right)^2}}$. 
Consequently, the Euclidean action can be expressed by the order parameter as
\begin{equation}\label{eq:kin3}
\begin{split}
    S_E=\beta\int d^3\vec{x}\left[\frac{9 T^2}{2-\frac{(3l-1)^2}{2}}\bm{\nabla}l\cdot\bm{\nabla}l+V_{PLM}(l)\right].
\end{split}
\end{equation}
Now, we have obtained the general form of the effective action, which includes the composite nature of the order parameter for the $SU(3)$ confinement phase transition at least at tree level. In comparison, the kinetic terms in Eq.~\eqref{eq:kin3} range from 4.5 to 6, which is 1.5 to 2 times larger than the value reported by Pisarski in~\cite{Dumitru:2000in}.

The above result can be generalized to an arbitrary color number. We first return to the kinetic terms expressed by $L$, $K = - T^2 \mathrm{Tr}\left[ \bm{\nabla} L^{-1} \cdot \bm{\nabla} L \right]$, where $L$ can be found for an arbitrary color number $N$ by replacing 3 in Eq.\eqref{lsu3} by $N$. So, in this case, the kinetic term can also be expressed by $L$ and $L^\dagger$, together with $N$,
\begin{equation}
    K=-N^2T^2\mathrm{Tr}\left(\bm{\nabla}L^\dagger\cdot\bm{\nabla}L\right).
\end{equation}
As we do in the $SU(3)$ case, the Polyakov loop matrix can be parametrized through the eigen phase $q_i$ as $\diag(e^{q_1},e^{q_2},\cdots,e^{q_N})$. Therefore, the trace of the product of gradients is
\begin{equation}\label{eq:trnlnl}
    \mathrm{Tr}\left(\bm{\nabla}L^\dagger\cdot\bm{\nabla}L\right)=\sum_i \bm{\nabla}q_i \cdot \bm{\nabla}q_i.
\end{equation}
Just following the same equal-eigenvalue condition for $N_c=3$ and generate them into the theory with $N_c>3$, we can express $q_i$ under the constraint $\sum q_i=0$ as $q_i =2\pi\left(-\frac{N+1}{2N}+\frac{i}{N}\right)q$, and then substituting it into Eq.\eqref{eq:trnlnl}, we find
\begin{equation}\label{eq:trnlnlqq}
\mathrm{Tr}\left(\bm{\nabla}L^\dagger\cdot\bm{\nabla}L\right) = \frac{\pi^2}{3}\frac{N^2-1}{N}\bm{\nabla}q\cdot\bm{\nabla}q.
\end{equation}
Finally, we can compute the order parameter $l$ following the definition, which yield $l=f(q)=\frac{1}{N}\frac{\sin(\pi q)}{\sin(\pi q/N)}$. As consequences, we find the kinetic for $SU(N_c)$ confinement phase transition as follow
\begin{equation}
K =Z_c(l)\bm{\nabla}l\cdot\bm{\nabla}l,
\end{equation}
where
\begin{equation}\label{nz}
Z_c(l)= -\frac{N^3(N^2-1)}{3\left[N\cot(\pi f^{-1}(l))-\cot\left(\frac{\pi f^{-1}(l)}{N}\right)\right]^2}\frac{T^2}{l^2}.
\end{equation}
In the above expression, $f^{-1}$ is the inverse function of $f$, which has no analytic form for a general color number, but in practical we can always find it through the numerical method. In the following, we retain $Z_c(l)$ and do not normalize the kinetic term into the canonical form by redefining the field~\cite{Hua:2025fap}, as we wish to preserve the intrinsic information of the order parameter.

\subsection{The Effect of the Non-trivial renormalization factor}
With the expressions of the kinetic terms, let us start to discuss how those kinetic terms affected the physical observation quantities: the nucleation rate and the corresponding GWs from the phase transitions. We would discuss this effect on three independent effective models: 1) The Haar-Measure PLM; 2) The Polyakov Polynomial PLM; 3) The quasi-particles Models. Let us first briefly introduce those models and the corresponding effective potential.

\subsubsection{The Haar-Measure and Polynomial PLM}
Begin with the Haar measure PLM, which was first constructed by Fukushima~\cite{Fukushima:2003fw, Fukushima:2017csk} to describe the $SU(3)$ QCD phase transition, and then developed by~\cite{Roessner:2006xn,Fukushima:2010bq} to describe the thermal dynamics. The effective potential in this model can be read as
\begin{equation}
    V_{Haar}(l,T)=T^4 \left[-\frac{1}{2}A_1(T)l^2+A_2(T)\log(1-6l^2+8l^3-3l^4)\right],
\end{equation}
with
\begin{align}
    A_1(T)=a_0+a_1\frac{T_0}{T}+a_2\left(\frac{T_0}{T}\right),\ \ A_2(T)=a_3\left(\frac{T_0}{T}\right),
\end{align}
where $T_0=0.27{\rm GeV}$, $a_0=3.51$, $a_1=-2.47$, $a_2=15.2$, $a_3=-1.75$ are constrained and fixed by imposing some physical conditions and fitting the effective potential to the lattice data on the thermodynamic quantities. This potential is constructed following the rules of nearest-neighbor coupling among PLs and the $Z_3$ center symmetry. It works perfectly when describing the $SU(3)$ thermodynamics~\cite{Roessner:2006xn}. 

However, the Haar-Measure PLM meets some troubles if we want to discuss the general $SU(N_c)$ theory for $N_c>3$. The numbers of dependent fields increase with the color numbers $N_c$, which makes the nucleation equation impossible to solve if $N_c$ is quite large~\cite{Kubo:2018vdw}. Moreover, the thermal dynamics generated by this model are inconsistent with $N_c^2-1$ scaling rules, which are provided by lattice results~\cite{Kang:2021epo}. As a result, we only imply this model when $N_c=3$.

To discuss the confinement phase transition for the general color number, one can focus on the Polynomial PLM, which is firstly constructed following the Landau-Ginzburg order parameters theory with the $Z_3$ center symmetry for QCD~\cite{Ratti:2006gh}. Different from the Haar-Measure PLM, the Polynomial PLM can be consistently generated into the general $SU(N_c)$ theory by considering the effective interaction operators between the order parameter~\cite{Pisarski:2001pe}. After fitting all the lattice data~\cite{Huang:2020crf,Kang:2021epo}, one can find the effective potential in these models as
\begin{equation}
\begin{split}
    &V_{Poly}(l,T)=T^4\left(-\frac{1}{2}B(T)l^2-2cl^3+dl^4+el^6+fl^8\right),
    \\
    &B(T)=b_0+b_1\frac{T_0}{T}+b_2\left(\frac{T_0}{T}\right)^2+b_3\left(\frac{T_0}{T}\right)^3+b_4\left(\frac{T_0}{T}\right)^4,
    \end{split}
\end{equation}
where the phenomenon parameters obtained by fitting the lattice data are listed in the Table.\ref{tablepolynomial}.

\begin{table}[t!]
\centering
\begin{tabular}{|c|c|c|c|c|c|c|c|c|c|c|c|}
\hline
 & \multicolumn{9}{c|}{polynomial potential} & \multicolumn{2}{c|}{quasi-particle model} \\
\hline
$N_c$ & $b_0$ & $b_1$ & $b_2$ & $b_3$ & $b_4$ & $c$ & $d$ & $e$ & $f$ & $g$ & $h$ \\
\hline
3 & 3.72 & $-5.73$ & 8.49 & $-9.29$ & 0.27 & 2.4 & 4.53 & --- & --- & 2.7499 & 5.7727 \\
\hline
4 & 9.51 & $-8.79$ & 10.1 & $-12.2$ & 0.489 & --- & $-2.46$ & 3.23 & --- &  2.7203 & 7.9951 \\
\hline
6 & 16.6 & $-47.4$ & 108 & $-147$ & 51.9 & --- & $-54.8$ & 97.3 & $-43.5$ & 2.7099 & 10.0965 \\
\hline
8 & 28.7 & $-69.8$ & 134 & $-180$ & 56.1 & --- & $-90.5$ & 157 & $-68.9$ & 2.7083 & 10.9954 \\
\hline
\end{tabular}
\caption{Parameters of the effective potential for different models and color numbers.}
\label{tablepolynomial}
\end{table}

\subsubsection{Quasi-Particle Model}
We would also like to discuss another type of model, the quasi-particles model, which is originate from the hard-thermal-loop perturbation theory~\cite{Andersen:1999fw,Andersen:2000zn,Andersen:2002ey}
. Rather than constructing the effective potential from the symmetry of the order parameters like PLMs, the potential from the quasi-particle model can be directly derived from the first-principles lagrangian~\cite{Sasaki:2012bi, Reinosa:2014ooa,vanEgmond:2021jyx,Kang:2022jbg}. The core concept of this approach is that dominant gluon interactions can be absorbed into an effective, temperature-dependent mass term~\cite{Kang:2022jbg}. This results in a thermal system of weakly interacting quasi-particles whose effective potential can be derived from the lagrangaian. With the help of this model, we can write down the effective potential for the general $SU(N_c)$ theory as an analytic form~\cite{Kang:2022jbg}
\begin{equation}
\begin{aligned}
V_{qpa}(l, T, N) 
&= -\frac{N^2}{2} \Bigg[
\frac{3 T^4}{\pi^2} \left( \frac{M_g(T, N)}{T} \right)^2 K_2\!\left( \frac{M_g(T, N)}{T} \right) l^\dagger l \\
&\quad + \frac{\pi^2 T^4}{45} \bigg( [-1 + f^{-1}(l)]^2 \{-1 - 2f^{-1}(l) + 2[f^{-1}(l)]^2\} 
\\
&\quad\quad\quad\quad\quad- \frac{5 [-1 + f^{-1}(l)]^2 [f^{-1}(l)]^2}{N^2} + \frac{[f^{-1}(l)]^3 [-4 + 3f^{-1}(l)]}{N^4}\bigg)\Bigg].
\end{aligned}
\end{equation}
Here, $K_2$ denotes the modified Bessel function of the second kind of order 2 and $f^{-1}(l)$ is the function defined in Eq.\eqref{nz}. Except from that, $M_g$ is the temperature-dependent quasi-particles mass, which can be formulated as
\begin{equation}
    M_{g}(T,N)=g T_0 - h (T - T_0).
\end{equation}
The parameters $g$ and $h$ are listed in Table.\ref{tablepolynomial}. With the mass parameters, this model has been proven that it can unified describe the confinement phase transition and the corresponding thermal dynamics down to $T_c$~\cite{Kang:2022jbg}, even with the weak quantum correction down to 2 loop~\cite{Reinosa:2015gxn}.

\subsubsection{The Effect of the Phase Transition and GWs}\label{confinementgws}
Having established the necessary elements, we proceed to compute the true vacuum nucleation rate and the resulting GWs spectrum. As outlined in Sec.~\ref{ACTION}, the thermal nucleation rate per unit volume is calculated from the $O(3)$-symmetric Euclidean action. For the confinement phase transition, this action is given by
\begin{equation}
S_3(T)=4\pi \int_0^{\infty} r^2dr \left[\frac{Z_c(l)}{2}\left(\frac{dl}{dr}\right)^2+V_{PLM}\right].
\end{equation}
Here, we define $Z_l$ as $1/Z_c(l)$ and $Z_l'=dZ_l/dl$, and the Euclidean action is evaluated by solving the corresponding tunneling equation together with the appropriate boundary conditions.
\begin{equation}\label{eomofl}
\begin{aligned}
    \frac{d^2l}{dr^2}+\frac{2}{r}\frac{dl}{dr}-\frac{1}{2}\frac{Z_l'}{Z_l}\left(\frac{dl}{dr}\right)=Z_l\frac{\partial V_{PLM}}{\partial l},\\
    \left.\frac{dl}{dr}\right|_{r=0}=0,\lim_{r\to\infty}l(r)=1.
\end{aligned}
\end{equation}
Then, the nucleation temperature $T_n$~\footnote{One should, in principle, use the percolation temperature $T_p$. However, since for the QCD phase transition, without strong supercooling, $T_n$ and $T_p$ are close. So, for convenience, we still use the $T_n$ to do the evaluation.} can be determined from the nucleation condition $S_3(T)/T \sim 140$~\cite{Quiros:1999jp,Kang:2020jeg}, after which the phase transition parameters can be calculated. This equation can be solved by using our numerical package \texttt{VacuumTunneling}~\cite{Hua:2025fap}, which is specially designed to solve the bounce equation with non-trivial renormaliztion factor.

The parameter characterizing the strength of the phase transition is denoted by $\alpha$, which is defined as
\begin{equation}
\alpha \equiv \frac{L_N(T_n)}{\rho_{\rm SM}(T^{\rm SM}_n) + \rho_{\rm DM}(T_n)},
\end{equation}
here, the latent heat of the phase transition, $L_N$, is the change in energy density caused by the transition at the nucleation temperature $T_n$:
\begin{equation}
L_N(T_n) = \Delta V_{eff}(T_n) - T \frac{\partial \Delta V_{eff}(T)}{\partial T} \Bigg|_{T=T_n},
\label{LN-def}
\end{equation}
and $T^{\mathrm{SM}}$ denotes the temperature of the Standard Model sector, which is proportional to the dark sector temperature and depends on the proportionality factor $\zeta$. As we work on the radiation-dominated era, the energy density $\rho_i$ can be expressed as $\rho_i = \frac{\pi^2}{30} g_i^* T_i^4$, where $i = 1$ or $2$ represents the Standard Model sector or the dark matter sector, respectively. $g^*$ is the number of relativistic degrees of freedom, $g^*$ for the dark sector is $4 N_c N_f + 2 (N_c^2 - 1)$.

The parameter $\beta$ characterizes the rapidity of the phase transition and is defined as
\begin{equation}
\begin{aligned}
\widetilde \beta=T_{n} \frac{d}{dT}\left(\frac{S_3(T)}{ T}\right)
\Bigg|_{T=T_{n}},
\label{betat}
\end{aligned}
\end{equation}
which is inversely proportional to the ratio of the phase transition duration to the Hubble time. If all the above quantities, has been computed, one can use them to compute the GWs.

\begin{table}[h!]
\centering
\begin{tabular}{|c|c|c|c|c|c|}
\hline
Model & $N_c$ & kinetic term & $T_n$ & $\alpha$ & $\beta$ \\ 
\hline
\multirow{2}{*}{Harr-measure}&\multirow{2}{*}{3} 
 & $Z=1$ & 0.26915 & 0.386167 & 107122 \\
 & & $Z\ne 1$ & 0.266789 & 0.346265 & 34416.7 \\
\hline
\multirow{8}{*}{Polynomial} 
 & \multirow{2}{*}{3} & $Z=1$ & 0.26914 & 0.399159 & 101566 \\
 & & $Z\ne 1$ & 0.266832 & 0.344917 & 39134.5 \\
\cline{2-6}
 & \multirow{2}{*}{4} & $Z=1$ & 0.268926 & 0.439018 & 149449 \\
 & & $Z\ne 1$ & 0.267057 & 0.384421 & 64252.7 \\
\cline{2-6}
 & \multirow{2}{*}{6} & $Z=1$ & 0.26654 & 0.522667 & 42223 \\
 & & $Z\ne 1$ & 0.256473 & 0.428721 & 9938.46 \\
\cline{2-6}
 & \multirow{2}{*}{8} & $Z=1$ & 0.26855 & 0.537504 & 52057.8 \\
 & & $Z\ne 1$ & 0.258188 & 0.445281 & 11473.5 \\
\hline
\multirow{8}{*}{Quasi-particle} 
 & \multirow{2}{*}{3} & $Z=1$ & 0.269796 & 0.292612 & 439020 \\
 & & $Z\ne 1$ & 0.269418 & 0.228584 & 409459 \\
\cline{2-6}
 & \multirow{2}{*}{4} & $Z=1$ & 0.269831 & 0.406925 & 516254 \\
 & & $Z\ne 1$ & 0.269434 & 0.30979 & 538459 \\
\cline{2-6}
 & \multirow{2}{*}{6} & $Z=1$ & 0.269881 & 0.492672 & 718533 \\
 & & $Z\ne 1$ & 0.269517 & 0.36447 & 725276 \\
\cline{2-6}
 & \multirow{2}{*}{8} & $Z=1$ & 0.269906 & 0.525231 & 881391 \\
 & & $Z\ne 1$ & 0.269556 & 0.36144 & 974773 \\
\hline
\end{tabular}
\caption{Phase transition parameters for different models, $N_c$, and $Z$ sectors}
\end{table}

Gravitational waves generated by a first-order cosmological phase transition originate from three main sources: the energy-momentum tensor from bubble wall collisions, sound waves in the plasma, and turbulence in the plasma. The energy of these three types of gravitational waves all originates from the latent heat of the phase transition, but the fraction of the released energy transferred to each source differs, and is characterized by an efficiency factor $\kappa$. First, consider the gravitational waves directly produced by bubble collisions. The energy fraction $\kappa_{col}$ is defined as the ratio of the bubble wall energy $E_{wall}$ to the total vacuum energy $E_V$:
\begin{equation}
\kappa_{col} \equiv \frac{E_{wall}}{E_V}.
\end{equation}
Since the phase transition strength in the model studied here is relatively small, the resulting $\kappa_{col} \ll 1$. Therefore, this contribution to the gravitational waves can be neglected, and only the gravitational waves sourced by the two fluid effects need to be considered.

Next, we consider the gravitational waves sourced by sound waves in the plasma, characterized by the energy fraction $\kappa_{ sw}$. It should ideally be expressed in terms of the effective phase transition strength, $\alpha_{eff} = \alpha (1 - \kappa_{col})$. However, since $\kappa_{col} \ll 1$, we can approximate it directly as
\begin{equation}
\kappa_{sw} \approx \frac{\alpha}{0.73 + 0.083 \sqrt{\alpha} + \alpha}.
\end{equation}
The peak frequency of the sound waves is obtained numerically and can be expressed as
\begin{equation}
f_{sw,*} = \frac{2 (8 \pi)^{1/3}}{\sqrt{3}(v_w - c_s), R_*},
\end{equation}
where $v_w$ is the bubble wall velocity and $c_s$ is the speed of sound in the plasma. In this work, we focus on the radiation-dominated era, so $c_s = 1/\sqrt{3}$. $R_*$ is the average bubble separation, which can be expressed in terms of the nucleation rate $\Gamma(T_n)$ at the nucleation temperature $T_n$ using an exponential approximation and the characteristic timescale $1/\beta_n$:
\begin{equation}
R_* = (8 \pi)^{1/3} \frac{v_w}{\beta_n}.
\end{equation}
The frequency given above corresponds to the peak frequency at the time of bubble nucleation ($t_n$). As the universe evolves to the present time ($t_0$), the gravitational waves will be redshifted. Therefore, the frequency received by detectors today should be multiplied by the ratio of scale factors $a(t_0)/a(T_n)$, giving the peak frequency of the sound waves as
\begin{equation} \label{} 
\begin{split}
f_{sw}&=1.65\times 10^{-5}\left(\frac{T_n}{100 \rm GeV}\right)\left(\frac{g_*}{100}\right)^{\frac{1}{6}} \frac{3.4}{(v_w-c_s)H_* R_*} \rm Hz
\\
&=2.75\times10^{-5}\frac{\widetilde\beta}{v_{w}}\left(\frac{T_n}{100 \rm GeV}\right)\left(\frac{g_*}{100}\right)^{\frac{1}{6}} \rm Hz.
\end{split}
\end{equation}
In addition, it is necessary to know the frequency spectrum and the amplitude of the gravitational waves. The frequency spectrum is given by the numerically fitted shape factor $S_{sw}(f)$:
\begin{align} 
S_{sw}(f)&=(f/f_{sw})^3\left[\frac{7}{4+3(f/f_{sw})^2}\right]^{\frac{7}{2}}.
\end{align}
The amplitude of the gravitational wave spectrum depends on the bubble parameters $H_*$, the lifetime of the sound waves $\tau_{sw}$, and the phase transition strength $\alpha$. The resulting gravitational wave energy spectrum can be expressed as
\begin{align} \label{sound}
h^2\Omega_{sw}(f)&=6.35\times 10^{-6}\left( H_* R_*\right) \left( H_* \tau_{sw} \right)  \left(\frac{ \kappa_{sw}\alpha}{1+\alpha}\right)^{2}\left(\frac{100}{g^*}\right)^{\frac{1}{3}}v_{w} S_{sw}(f).
\end{align}
It should be noted that the lifetime parameter $\tau_{sw}$ needs to be chosen case by case. Generally, it can be compared with the Hubble time, but when the phase transition proceeds sufficiently quickly, the lifetime of the sound waves will correspondingly shorten. Thus, the value of $\tau_{sw}$ is given by
\begin{equation}
H_*\tau_{sw}=\min[1,H_* R_*/U_f],
\end{equation}
where $U_f$ in the denominator is the root-mean-square velocity of the fluid, implying that when the phase transition occurs rapidly, the bubble collision rate increases with $U_f$, leading to a shorter overall duration. For the phase transition model studied in this work, the timescale parameter $\beta$ is very large. In this case, the expression for $U_f$
\begin{equation} 
U_f\simeq \frac{\sqrt{3}}{2}\left(\frac{\alpha(1-\kappa_{col})}{1+\alpha(1-\kappa_{col})}\kappa_{sw}\right)^{1/2},
\end{equation}
can be substituted into $H_*\tau_{sw}=H_* R_*/U_f$, leading to the gravitational wave spectrum for rapidly occurring phase transitions:
\begin{align} \label{soundGW}
h^2\Omega_{sw}(f)&=6.3\times 10^{-5}\frac{1}{\widetilde\beta^2}\left(\frac{ \kappa_{sw}\alpha}{1+\alpha}\right)^{2}\left(\frac{100}{g^*}\right)^{\frac{1}{3}}v_{w}^2 S_{sw}(f).
\end{align}

After the propagation of the plasma sound waves ends, a portion of the energy will be transferred into turbulence. If the lifetime of the sound waves can last longer than the Hubble time $1/H$, most of the energy will go into the sound waves, leaving only a small fraction for turbulence. The energy fraction of turbulence is roughly $\kappa_{turb}\sim 0.05\kappa_{sw}$. In this case, some studies provide the gravitational wave spectrum from turbulence as
\begin{equation}\label{TURB}
h^2\Omega_{turb}(f)=3.3\times10^{-4}\frac{1}{\widetilde\beta}\left( 1-\frac{2(8\pi)^{1/3} v_w}{\sqrt{3}\widetilde\beta}\right)\left(\frac{ \kappa_{sw}\alpha}{1+\alpha}\right)^\frac{3}{2}\left(\frac{100}{g}\right)^{\frac{1}{3}}v_{w} S_{turb}(f),
\end{equation}
including the shape factor
\begin{equation} 
\begin{split}
S_{turb}(f)=\frac{(f/f_{turb})^3}{[1+(f/f_{turb})]^{\frac{11}{3}}(1+8\pi f/h)}.
\end{split}
\end{equation}
In this study, the duration of the phase transition sound waves is very short. Although this may lead to more energy being transferred into turbulence, the highly nonlinear nature of turbulent motion also converts part of the sound wave energy into heat. Therefore, studying the gravitational waves from turbulence is very challenging. Therefore, in this work, a fraction of 0.05 is still adopted as the energy fraction for turbulence, and the gravitational waves sourced by turbulence are calculated using the above spectral formula and shape factor.

Another important parameter entering the gravitational wave spectra is the wall velocity $v_w$, which characterizes the speed of bubble expansion~\cite{Ramsey-Musolf:2025jyk}. The lower the wall velocity, the more easily the thermal bath absorbs the energy released by the phase transition, resulting in a smaller gravitational wave amplitude. Conversely, a higher wall velocity transfers a larger fraction of the energy into gravitational-wave production, leading to a larger amplitude. However, $v_w$ is a quantity difficult to precisely determine. Therefore, we take $v_w=1$ to provide an upper bound on the gravitational wave amplitude, and $v_w=0.1$ to give an estimate corresponding to a low wall velocity~\cite{Kang:2024xqk}.

\begin{figure}
    \centering
    \includegraphics[width=0.49\linewidth]{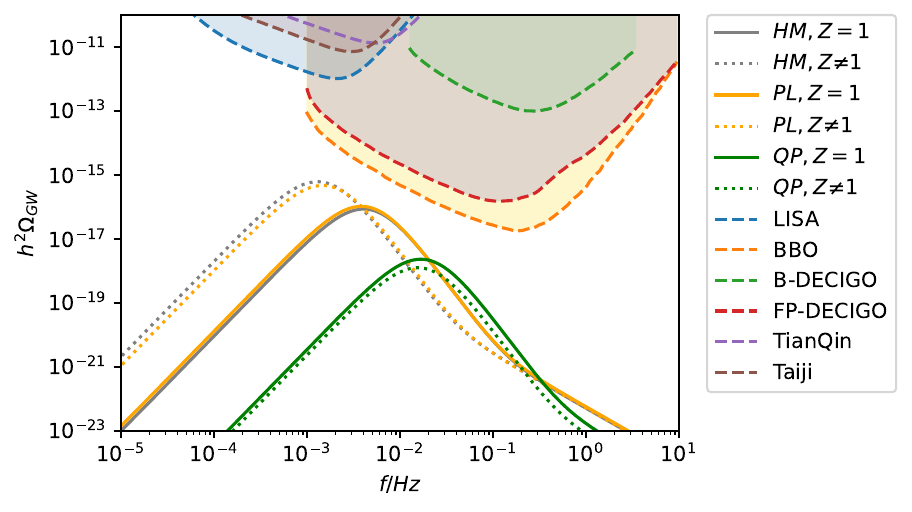}
    \includegraphics[width=0.49\linewidth]{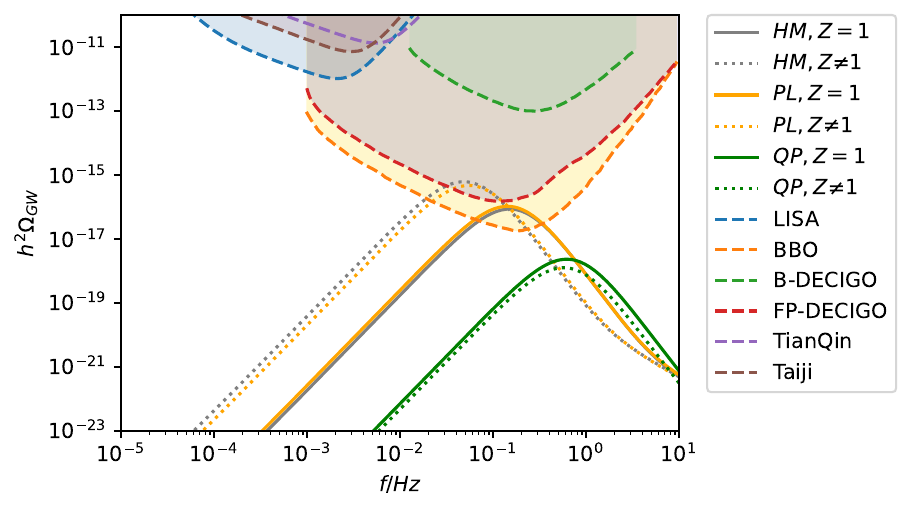}
    \caption{Gravitational wave spectra from different models for $N_c=3$, The left panel corresponds to $T_0 = 0.27,\mathrm{GeV}$, while the right panel corresponds to $T_0 = 10,\mathrm{GeV}$. Here, HM, PL, and QP denote the Haar measure, polynomial, and quasi-particle models, respectively. Solid lines represent the trivial kinetic term, and dotted lines the nontrivial one.}
    \label{3modelgws}
\end{figure}

\begin{figure} 
    \centering
    \includegraphics[width=0.49\linewidth]{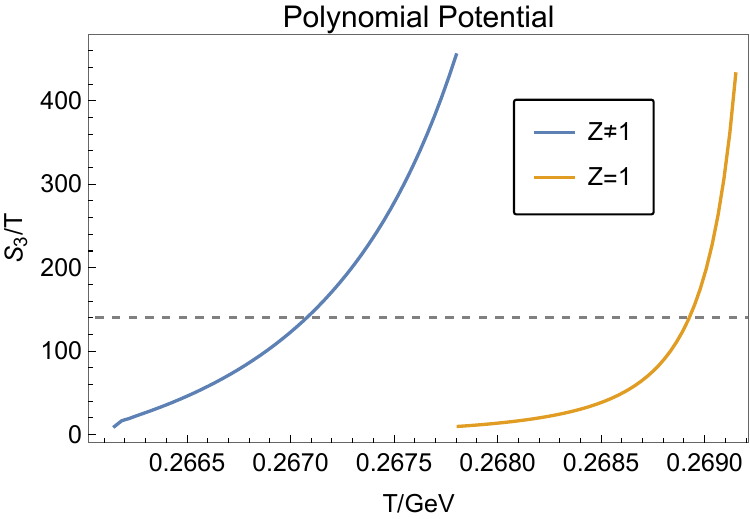}
    \includegraphics[width=0.49\linewidth]{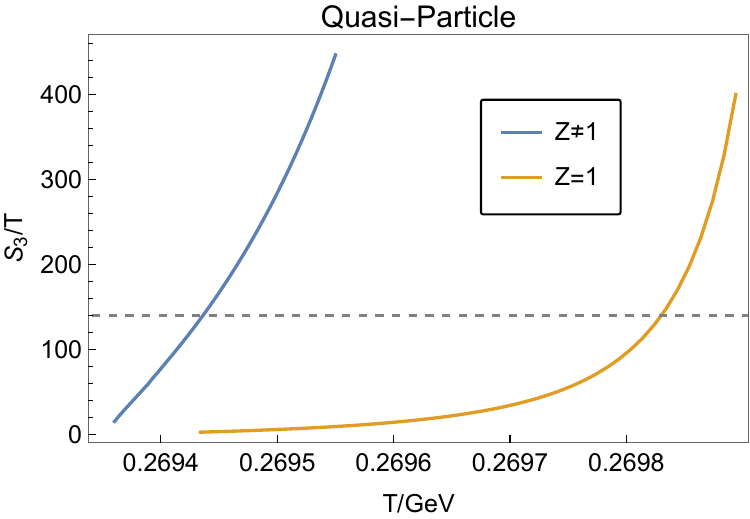}
    \caption{Temperature dependence of $S_3/T$. The blue curve shows the case with non-trivial kinetic terms ($Z\neq1$), while the orange curve shows the case with only the trivial kinetic term ($Z=1$). The dashed line indicates $S_3/T=140$. The left panel corresponds to the Polynomial Potential, and the right panel corresponds to the Quasi-Particle model.}
    \label{s3otot}
\end{figure}

\begin{figure}[htbp]
	\centering
    \begin{tikzpicture}
	    \draw [->] (-1,0)--(8,0) node[below right] { $\sigma$ };
	    \draw [->] (0,-1.7)--(0,2.5) node[above left] { $-V_{eff}$ };
		\draw[blue,line width=1.5pt,domain=0:7,samples=100] plot(\x,{ -\x^4 / 37.5 +\x^3 /3 -\x^2});
		\draw[fill=red] (0,0) circle (2.5pt) node[above left]{$\sigma_F$};
		\draw[dashed,brown,line width=1.5pt,domain=0:7,samples=100] plot(\x,{ -\x^4 / 75 +\x^3 /6 -\x^2 /2});
		\draw (7,3.3) ;
        \draw[fill=red] (6.48,1.69) circle (2.5pt) ;
        \draw (6.48,-0.5) node{$\sigma_T$};
        \draw[dash dot,gray,line width=1.5pt] (6.48,0)--(6.48,1.6);
        \draw[fill=red] (6.48,0.845) circle (2.5pt) ;
        \draw[fill=red] (5.5,0.81) circle (2.5pt) node[above left]{$\sigma_i$};
        \draw [->,line width=1pt] (5.35,0.81)--(5.1,0.4);
	\end{tikzpicture} 
	\caption{The blue solid line is the effective potential, in which shooting without renomalization occurs. The brown dashed line is a new potential transformed by $Z_\sigma$. The two lines have the same positions of extrema.}
    \label{1dshooting}
\end{figure}
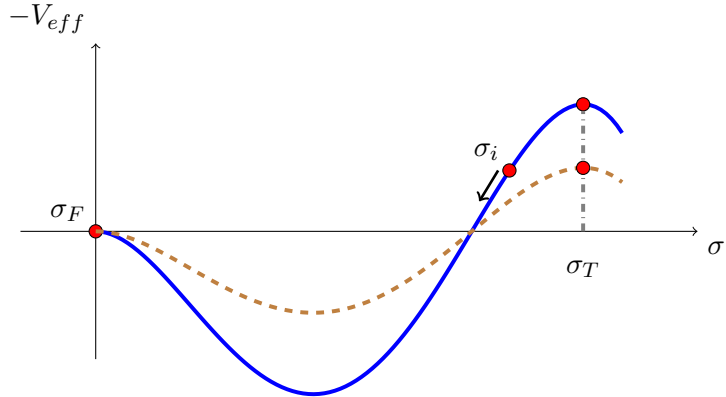

The gravitational wave spectra of different models for $N_c=3$ are shown in Fig.\ref{3modelgws}, together with the sensitivity curves of LISA~\cite{Caprini:2015zlo,LISA:2017pwj,Caprini:2019egz}, BBO~\cite{Crowder:2005nr,Corbin:2005ny,Harry:2006fi}, DECIGO~\cite{Yagi:2011wg,Seto:2001qf,Isoyama:2018rjb}, TianQin~\cite{TianQin:2015yph,TianQin:2020hid}, and Taiji~\cite{Ruan:2018tsw,Ruan:2020smc}. As seen in the figure, the gravitational wave amplitude of the Haar-measure potential and the ploynomial potential increases while that of quasi-particle model decreases when the non-trivial kinetic term is taken into account, compared to the case with the trivial one. Since the renormalization factor varies relatively slowly here, we can simply understand its effect as equivalent to multiplying the effective potential in the tunneling equation by a coefficient smaller than 1, as shown in Fig.\ref{1dshooting}. This effectively reduces the barrier height of the effective potential, but also decreases the potential difference between the true and false vacua. As a result, tunneling becomes more difficult to occur, requiring the temperature to drop further to satisfy the nucleation condition. As the temperature decreases, $S_3/T$ first drops rapidly and then slowly approaches zero. For $N_c=3$, after including $Z_l$, all three models have the nucleation temperature located in the region where the change of $S_3/T$ is slower, as shown in the left panel of Fig.~\ref{s3otot}, leading to a smaller $\beta$. However, in the lower-temperature regime, the effective potential changes more slowly, so a lower nucleation temperature corresponds to a weaker phase transition strength. For the Haar-measure potential and the polynomial potential, the change in $\beta$ is substantial, and dominates over the suppression of the gravitational wave amplitude coming from the decrease in $\alpha$. As a result, the gravitational wave amplitude becomes significantly larger. In contrast, for the quasi-particle model, the reduction in $\beta$ is not sufficient to compensate for the reduction in $\alpha$, so the gravitational wave amplitude becomes smaller.

\begin{figure}
    \centering
    \includegraphics[width=0.49\linewidth]{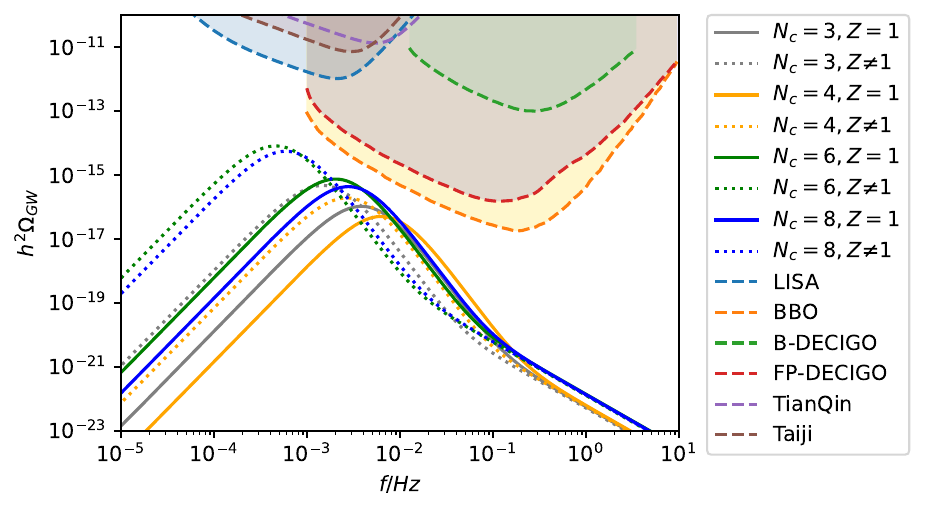}
    \includegraphics[width=0.49\linewidth]{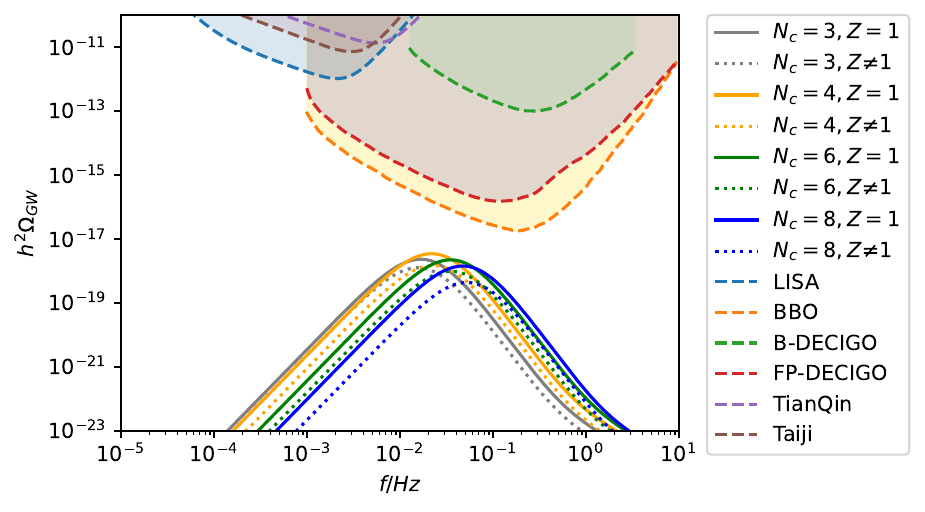}
    \caption{The left panel shows gravitational wave spectra from polynomial potential model, and the right panel shows that from quasi-particle model, both for $N_c=3,4,6,8$. Solid lines represent the trivial kinetic term, and dotted lines the nontrivial one.}
    \label{plgws}
\end{figure}
For $N_c=3,4,6,8$, we calculated the gravitational wave spectra from the polynomial potential and quasi-particle model, shown in Fig.\ref{plgws}. For $N_c=4,6,8$, the variation of $S_3/T$ in the polynomial potential is shown in the left panel of Fig.~\ref{s3otot}. After including the non-trivial kinetic terms, the point where $S_3/T\sim140$ lies in a flatter temperature region compared to the case without including them, similar to the $N_c=3$ situation discussed in the previous paragraph. In contrast, for the quasi-particle model, when only the trivial kinetic term is considered, the point where $S_3/T\sim140$ has already entered the flat region of the temperature dependence, whereas after including the non-trivial kinetic terms, the point where $S_3/T\sim140$ is still located in the rapidly varying region, as in the right panel of Fig.~\ref{s3otot}. Therefore, including the non-trivial kinetic terms decreases $\beta$ for the polynomial potential, but increases $\beta$ for the quasi-particle model. As a result, the non-trivial kinetic terms enhance the gravitational wave amplitude for the polynomial potential, while they suppress the gravitational wave amplitude for the quasi-particle model.

\section{The effect of the PL kinetic terms inside the chiral phase transition}\label{LAG}
With the bosonic part fixed, let us now discuss the fermionic contributions. After adding back quarks, QCD admits two types of phase transitions: 1) The confinement phase transition; 2) The chiral phase transition. Unlike the pure Yang-Mills sector, the Polyakov loops only serve as the quasi-order parameters of the confinement phase transitions because the $Z_N$ symmetry is no longer preserved, which causes $l$ to lose some nice properties, which would cause some troubles. We will introduce those problems later. The chiral phase transition, on the other hand, can be described by the symmetry breaking of chiral symmetry $SU(3)_L\otimes U(1)_V\otimes SU(3)_R\otimes U(1)_A$, whose order parameters are the chiral condensate $\langle \bar{q}q\rangle$. We need to use the effective theories as a bridge to construct the $\langle \bar{q}q\rangle$ from the fermion sectors, due to strong dynamics. To do that, we would like to use the NJL models.

\subsection{Zero-temperature effective potential}\label{vfermion}
To fulfill our goals, we assume that the fermion sector shares certain similarities with real QCD with 3 types of quark(u,d,s). We construct the model through the chiral symmetry and then the NJL Lagrangian reads
\begin{equation}\label{chrial}
\begin{split}
    \mathcal{L}_{\rm chiral}=\bar{q}i\bar{\slashed D} q+G_S\sum_{a=0}^8\left[(\bar{q}\lambda^a q)^2+(\bar{q}i\gamma^5\lambda^a q)^2\right]
    +G_D[\det(\bar{q}(1-\gamma^5)q+\bar{q}(1+\gamma^5)q)],
\end{split}
\end{equation}
where $q$ represents $u$, $d$ or $s$, $\lambda^a$ are the generators of $SU(3)$, normalized to 2 and $\bar{D}_\mu=\partial_\mu+iA_0\delta_{\mu 0}$ is the covariant derivative of the Polyakov background field. The detailed derivation of the effective potential is presented in \cite{Kang:2025nhe}; here we simply give its explicit expression. Following the idea of the NJL model, we define the quark condensate field as
\begin{equation}
\sigma = 4G_S\langle\bar{q}q\rangle,
\end{equation}
then, the NJL Lagrangian \eqref{chrial} can be written as
\begin{equation}\label{sigmaNJL}
    \mathcal{L}^{eff}_{NJL}=\bar{q}_i[i\slashed\partial-M_i(\sigma)]q_i-V_{NJL}^{tree}(\sigma),
\end{equation}
where
\begin{equation}\label{treep}
\begin{split}
    \mathcal{V}_{NJL}^{tree}(\sigma)&=\frac{3\sigma^2}{8G_S}+\frac{G_D\sigma^3}{128G_S^3},
\end{split}
\end{equation}
is the tree-level potential. The cross term between the condensate field and the fluctuation field gives rise to the quark mass term,
\begin{equation}
    M_i(\sigma)=-\sigma-\frac{G_D\sigma^2}{64G_S^2},
\end{equation}
which is the same for quarks of all flavors.

In addition to $\mathcal{V}_{NJL}^{tree}$, the vacuum energy $\mathcal{V}_{NJL}^{vac}$ also contributes to the zero-temperature effective potential. According to \cite{Kang:2025nhe}, one can obtain the vacuum energy as
\begin{equation}
    \mathcal{V}_{NJL}^{vac}=-N_c N_f\frac{\Lambda^4}{8\pi^2}\bigg[(2+\zeta^2)\sqrt{1+\zeta^2}+\frac{\zeta^4}{2}\log\left(\frac{\sqrt{1+\zeta^2}-1}{\sqrt{1+\zeta^2}+1}\right)\bigg],
\end{equation}
here, $\zeta=M(\sigma)/\Lambda$, where $\Lambda$ is the UV cut off for the NJL effective theory. In real QCD, $\Lambda$ can be determined from the meson spectrum. In dark QCD, however, there is no direct reference value for $\Lambda$. By requiring all energy scales to lie below $\Lambda$, it can be constrained to the range $\pi^2/(2N_c) \leq G_S \Lambda^2 \leq 3$ ~\cite{Kang:2025nhe}. 

\subsection{Finite-temperature effective potential}
Finite-temperature contributions from the quantum fields are present, analogous to the pure Yang-Mills case. The free energy is derived from the path integral of the Lagrangian in Eq.~\eqref{chrial}. A standard approach to this derivation is to work in the background Polyakov gauge $A_4$ and integrate out the fermion fields. This yields an effective potential for the interaction term of the form~\cite{Kang:2025nhe}:
\begin{equation}\label{vnjl}
\begin{split}
    V_{NJL}(\sigma,T)&=-2N_cN_f\int\frac{d^3\vec{k}}{(2\pi)^3}E_k
    \\
    &-2TN_f{\rm tr_c}\int\frac{d^3\vec{k}}{(2\pi)^3}\Bigg(\log\Bigg[ 1+e^{-\frac{(E_k-\mu+iA_4)}{T}}\Bigg]+\log\Bigg[ 1+e^{-\frac{(E_k+\mu-iA_4)}{T}}\Bigg]\Bigg).
\end{split}
\end{equation}
Let us focus on the case that $N_c=3$, by using Eq.~\eqref{define:Pol}, \eqref{lsu3} and Eq.~\eqref{matex}, the background gauge field can be rewritten in terms of the trace of the Polyakov loop~\cite{Kang:2025nhe}:
\begin{equation}
\begin{split}
    \mathcal{V}_{NJL}^{T}=-\frac{3T^4}{\pi^2}\int dx x^2 G(x,\sigma,T)
\end{split}
\end{equation}
where $G(x, \sigma, T)$ is given by
\begin{equation}
\begin{split}
    G(x,\sigma,T)&=
    \log\Big[1+e^{-3\sqrt{x^2+\frac{M(\sigma)^2}{T^2}}}+3l e^{-\sqrt{x^2+\frac{M(\sigma)^2}{T^2}}}+3l^*e^{-2\sqrt{x^2+\frac{M(\sigma)^2}{T^2}}}\Big]
    \\
    &+\log\Big[1+e^{-3\sqrt{x^2+\frac{M(\sigma)^2}{T^2}}}+3l^*e^{-\sqrt{x^2+\frac{M(\sigma)^2}{T^2}}}+3le^{-2\sqrt{x^2+\frac{M(\sigma)^2}{T^2}}}\Big].
\end{split}
\end{equation}

The final pieces of the effective potential come from the pure gluonic sector at finite temperature, which is exactly what we have discussed in Section.~\ref{Conf}. The form of this effective potential depends on which model is used to do the discussion. For now, we can label the pure gluonic contribution as $V_{model}$ where the index $model$ can be chosen as $Haar/Poly/qpa$. As a result, the total finite temperature effective potential can be written as 
\begin{equation}\label{PNJL}
    V_{eff}(\sigma,l,T)=V_{NJL}^{tree}(\sigma)+V_{NJL}^{vac}(\sigma)+V_{NJL}^T(\sigma,l,T)+V_{model}^T(l,T)
\end{equation}
However, as shown in Eq.\eqref{PNJL}, the fermion free energy would also contribute an effective potential which contains the Polyakov loops. This term would break the $Z_{N_c}$ center symmetry and cause some problems, as we mentioned at the beginning of this section. To be specific, in the most general case, in the confinement phase, $l\rightarrow0$ is still valid, but, in the confinement phase, $l\rightarrow1$ no longer always exists~\cite{Reichert:2021cvs}. To maintain these properties, one needs to refit the pure Yang-Mills models with the lattice QCD data. However, in our practice, we find that at the deconfinement phase with $T$ sufficiently large, $l$ is very close but slightly larger than 1, which would not change the main result. For convenience, we would still use the same phenomenon parameter in the gluon effective potential from the pure Yang-Mills case.

\subsection{Renormalization factor for quark condensate}

In principle, when studying the chiral phase transition under the influence of confinement, one should use the full effective action containing the kinetic terms of both $\sigma$ and $l$. Moreover, since $\sigma$ is a composite field and does not possess a tree-level kinetic term, its kinetic term must arise from loop corrections. Therefore, the action is given by
\begin{equation}\label{totalaction}
    S_E=\int d^3\vec{x}\left[\frac{1}{2}(Z_\sigma^{-1}\bm{\nabla}\sigma\cdot\bm{\nabla}\sigma+Z_l^{-1}\bm{\nabla}l\cdot\bm{\nabla}l)+V_{PNJL}\right].
\end{equation}
Here, $Z_\sigma$ is defined analogously to $Z_l$ in Sec.\ref{confinementgws} as the inverse of the coefficient of the kinetic term. The expression for the renormalization factor can be calculated as in Append.\ref{appendixd}, written as
\begin{equation}
    Z_\sigma^{-1}=2N_f N_c (1+\frac{G_D\sigma}{32G_S^2})^2[I(0,\sigma)+4M^2(\sigma)\frac{dI(0,\sigma)}{d\vec{q}^2}],
\end{equation}
here, the loop integral $I(0,\sigma)$ and $I'(0)$ is expressed as
\begin{align}
    I(0,\sigma)=\int\frac{d^3\vec{k}}{(2\pi)^3}\frac{1}{2E_k^2}\Bigg[\frac{\partial f(E_k,r)}{\partial E_k}+\frac{1-2f(E_k,r)}{2E_k}\Bigg],
    \\
    \frac{dI(0,\sigma)}{d\vec{q}^2}=\int\frac{d^3\vec{k}}{(2\pi)^{3}}\frac{1}{16E_{k}^{5}}\left[6f(\omega)-2E_{k}\frac{\partial f(\omega,r)}{\partial \omega}-1\right],
\end{align}
where 
\begin{equation}
    f(\omega,r).=\frac{1}{N_c}\sum_{j=1}^{N_c}\frac{1}{1+e^{\frac{\omega}{T}}e^{i2\pi q_j}}.
\end{equation}

\begin{figure}
    \centering
    \begin{minipage}[c]{0.49\textwidth}
        \includegraphics[width=\linewidth]{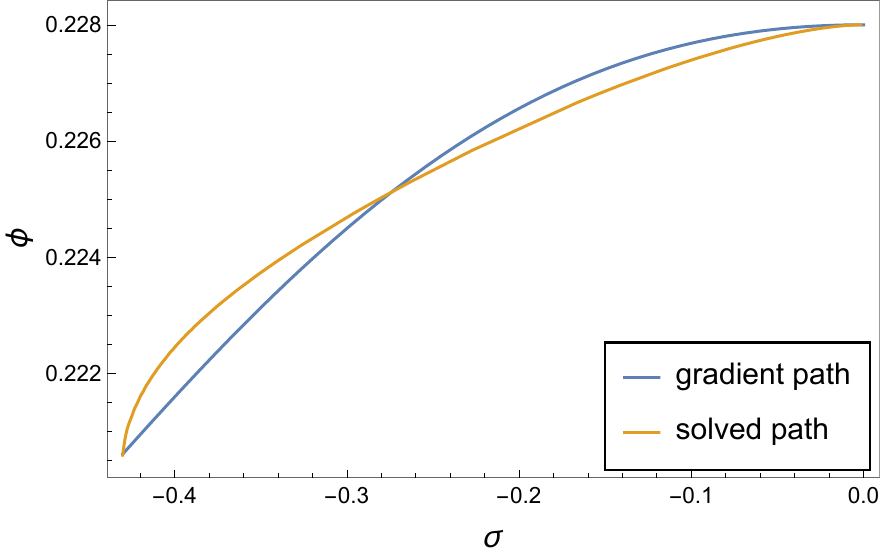}
    \end{minipage}
    \begin{minipage}[c]{0.49\textwidth}
        \begin{tabularx}{0.95\textwidth}{|c|X|X|c|c|}
            \hline
            Path & \multicolumn{2}{c|}{Kinetic Term} & $S_3/T$ & Comparison \\
            \hline
            Gradient & \centering $Z_\sigma$ & \centering --- & 140 & $100\%$ \\
            \hline
            Solved & \centering $Z_\sigma$ & \centering --- & 139.346 & $99.53\%$\\
            \hline
            Solved & \centering $Z_\sigma $ & \centering $ Z_l$ & 143.497 & $102.50\%$\\
            \hline
        \end{tabularx}
    \end{minipage}
    \caption{The left panel shows the path obtained by solving the equations of motion, along with the path in the gradient direction. Here, $\phi$ is defined by introducing the temperature scale as $\phi=lT$. The right panel shows the corresponding $S_3/T$, including the cases where the kinetic term of $l$ is included and where it is neglected for the solved path. This is computed under the Haar-measure model for the PL effective potential, with the parameters taken as $G_S = 2.2$ and $G_D = -282$.}
    \label{paths}
\end{figure}

\begin{figure}
    \centering
    \includegraphics[width=0.49\linewidth]{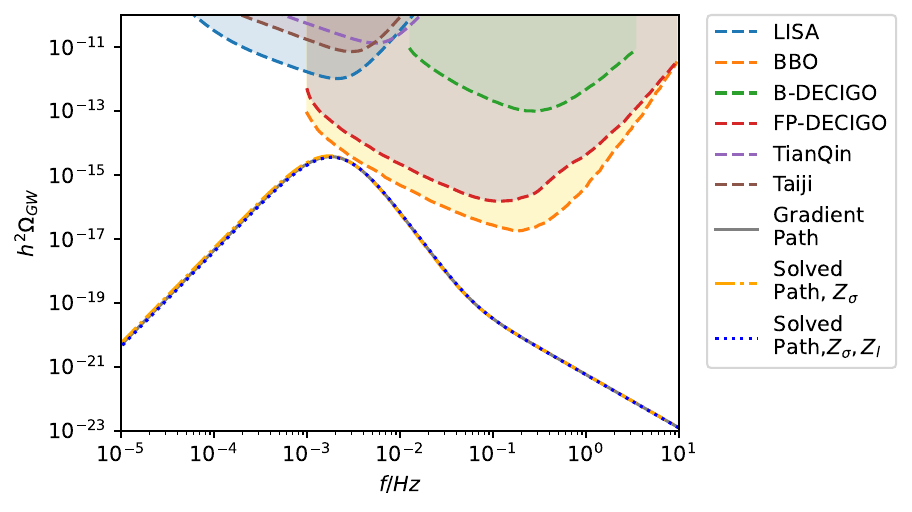}
    \includegraphics[width=0.49\linewidth]{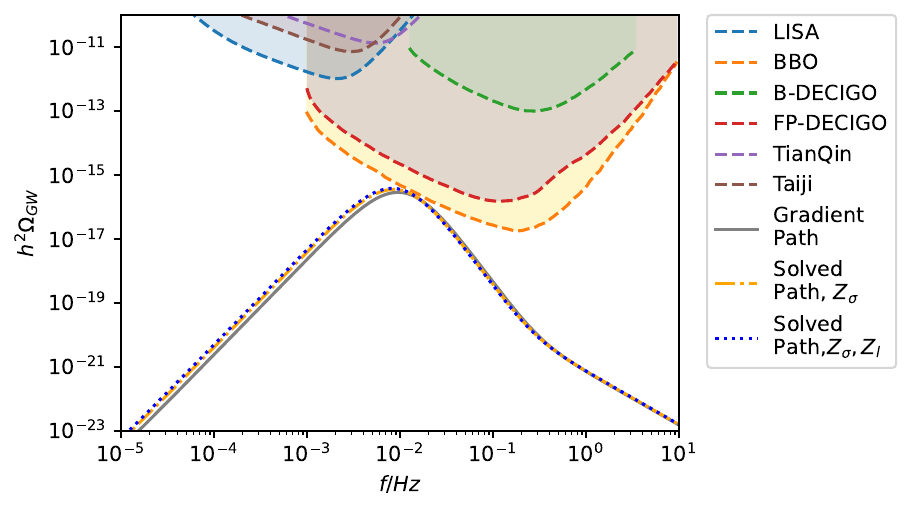}
    \includegraphics[width=0.49\linewidth]{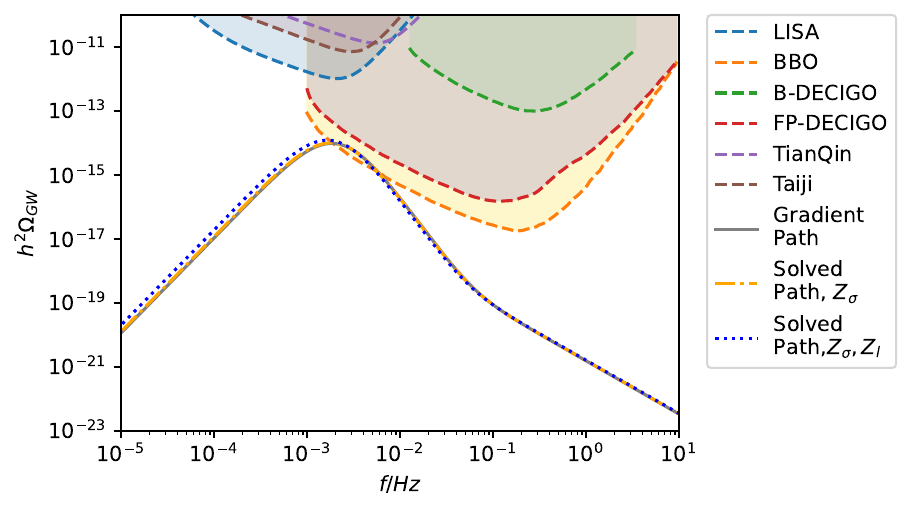}
    \includegraphics[width=0.49\linewidth]{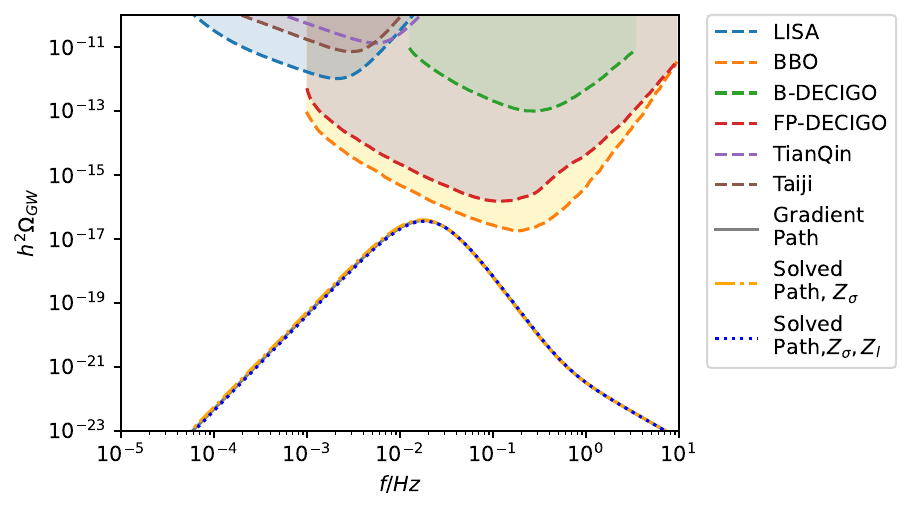}
    \caption{Gravitational wave spectra and detector sensitivity curves for the gradient path and the solved path, with and without the kinetic term of $l$. The upper-left panel corresponds to $G_S = 2.2$ and $G_D = -282$. The upper-right, lower-left, and lower-right panels vary one parameter relative to the upper-left case, with $G_S = 3.46$, $G_D = -481$, and $v_w = 0.1$, respectively. The PL effective potential is taken to be the Haar-measure model.}
    \label{gwpaths}
\end{figure}

\subsection{Numerical Solution of the Bounce Equation and Calculation of Chiral Phase Transition Parameters and the Corresponding GWs}\label{secivd}

\begin{table}
\centering
\begin{tabular}{|c|c|c|c|c|c|c|c|}
\hline
$N_c$ & $G_S$ & $G_D$ & PL Model & $T_c$ & $T_n$ & $\alpha$ (for $\zeta=0/1$) & $\beta$ \\ 
\hline
\multirow{15}{*}{3}
 & \multirow{3}{*}{2.2} & \multirow{3}{*}{-282} & Haar-measure & 0.2679 & 0.266192 & 0.985007/0.301296 & 39358 \\
 &                      &                      & Polynomial   & 0.25798 & 0.256266 & 1.15847/0.354356 & 38246.5 \\
 &                      &                      & Quasi-particle & 0.255975 & 0.254285 & 1.07840/0.329863 & 38412.6 \\
\cline{2-8}
 & \multirow{3}{*}{2.73} & \multirow{3}{*}{-282} & Haar-measure & 0.32407 & 0.32318 & 0.910041/0.278365 &  91722.5 \\
 &                      &                      & Polynomial   & 0.31445 & 0.313482 & 1.09300/0.334330 & 83212.8 \\
 &                      &                      & Quasi-particle & 0.314752 & 0.313864 & 0.963885/0.294835 & 91159.4 \\
\cline{2-8}
 & \multirow{3}{*}{3.46} & \multirow{3}{*}{-282} & Haar-measure & 0.37072 & 0.370089 & 1.04047/0.318261 & 149659 \\
 &                      &                      & Polynomial   & 0.361805 & 0.361024 & 1.26513/0.386982 & 119149 \\
 &                      &                      & Quasi-particle & 0.363027 & 0.362385 & 1.08513/0.331923 & 145105 \\
\cline{2-8}
 & \multirow{3}{*}{2.2} & \multirow{3}{*}{-481} & Haar-measure & 0.31716 & 0.314806 & 1.47026/0.449727 & 34193.2 \\
 &                      &                      & Polynomial   & 0.30878 & 0.306352 & 1.64114/0.501995 & 32367.6 \\
 &                      &                      & Quasi-particle & 0.307737 & 0.305370 & 1.56026/0.477257 & 33286.5 \\
\cline{2-8}
 & \multirow{3}{*}{2.2} & \multirow{3}{*}{-681} & Haar-measure & 0.355204 & 0.352528 & 1.81211/0.554293 & 34090.2 \\
 &                      &                      & Polynomial   & 0.347982 & 0.345156 & 1.95984/0.599480 & 31750.9 \\
 &                      &                      & Quasi-particle & 0.347248 & 0.344516 & 1.89290/0.579004 & 32849.8 \\
\hline
\multirow{2}{*}{4}
 & \multirow{2}{*}{2.2} & \multirow{2}{*}{-282} & Polynomial & 0.363551 & 0.361662 & 1.48368/0.590444 & 48975.3 \\
 &                      &                      & Quasi-particle & 0.362170 & 0.360349 & 1.45758/0.580058 & 50486.7\\
\hline
\multirow{2}{*}{6}
& \multirow{2}{*}{2.2} & \multirow{2}{*}{-282} & Polynomial & 0.499496 & 0.495422 & 1.49506/0.816533 & 31799.2 \\
 &                      &                      & Quasi-particle & 0.478221 & 0.474006 & 1.77365/0.968684 & 28861.4 \\
\hline
\end{tabular}
\caption{Phase transition parameters calculated under different model settings. The varied parameters include the number of colors ($N_c=3,4,6$), whether the universe is dominated solely by dark matter or jointly by dark matter and the Standard Model ($\zeta=0,1$), and the coupling constants ($G_S$, $G_D$). The listed quantities are the critical and nucleation temperatures ($T_c$, $T_n$), the phase transition strength ($\alpha$), and the inverse duration parameter ($\widetilde{\beta}$). For $N_c=3$, results are shown for the Haar measure potential, polynomial potential, and quasi-particle model, while for $N_c=4$ and $6$, only the quasi-particle and polynomial model are considered.}
\label{tablealphabeta}
\end{table}

Next, we examine how significant the impact of the kinetic term of $l$ is on the gravitational waves from the phase transition, and compare this effect with that induced by varying the model parameters. Just like the confinement phase transition, to determine the nucleation temperature, we adopt the nucleation condition $S_3(T_n)/T_n \sim 140$. We are still using our numerical package \texttt{VacuumTunneling}~\cite{Hua:2025fap} to obtain the 3-dimensional Euclidean action. 

As an illustration, we solved the equations of motion derived from the action in Eq.~\eqref{totalaction} within the Haar-measure model. The detailed expression is shown in Append.~\ref{appc}, because the expression is too complicated. The resulting path is found to be close to the one following the gradient direction of the effective potential, as shown in Fig.~\ref{paths}. Since $l$ is dimensionless, its dimension is introduced through $T$. During the chiral phase transition, the variation of $l$ is relatively small. As a consequence, the phase transition is dominated by the $\sigma$ variation. So the tunneling path can be approximately taken along the gradient direction of the effective potential, allowing the problem to be treated as a single-field problem of $\sigma$. The action of the complete solution differs by only about $2.5\%$ from the action obtained by solving the $\sigma$ single-field tunneling along the gradient direction of the effective potential, mainly due to the inclusion of the kinetic term of $l$. As a result, the choice of the path and the inclusion of the kinetic term of $l$ have only a minor effect on the gravitational-wave spectrum, with the change in the amplitude being much less than an order of magnitude, as shown in Fig.~\ref{gwpaths}.

To put the above conclusion into perspective, we now examine how the gravitational-wave spectrum varies with changes in the intrinsic model parameters, such as $G_S$, $G_D$, and $N_c$, and compare them with the effect discussed above.
Since we have shown that the kinetic term of $l$ has only a minor effect on the effective action, we use the following simplified effective action,
\begin{equation}\label{action31}
        S_E=4\pi\int d\rho\,\rho^2\left(\frac{1}{2Z_\sigma}\bigg(\frac{d\sigma}{d\rho}\bigg)^2+V_{PNJL}[\sigma,l(\sigma),T]\right),
\end{equation}
to compute the phase transition parameters. Thus, the corresponding equation of motion and the boundary conditions at the escape point $\rho = 0$ and the false vacuum $\rho \to \infty$ become:
\begin{align}\label{bounce}
  &  \frac{d^2\sigma}{d\rho^2}+\frac{2}{\rho}\frac{d\sigma}{d\rho}-\frac{1}{2}\frac{Z_\sigma'}{Z_\sigma}\left(\frac{d\sigma}{d\rho}\right)^2=Z_\sigma V_{PNJL}',\\
   & \frac{d\sigma}{d\rho}\bigg|_{\rho=0}=0, \quad \sigma|_{\rho\rightarrow\infty}=0,
\end{align}
here, the prime denotes derivative of $\sigma$. The resulting phase transition parameters are listed in Table~\ref{tablealphabeta}, after which the gravitational-wave spectrum can be computed following the procedure described in Sec.~\ref{confinementgws}.

\begin{figure}
    \centering
    \includegraphics[width=0.49\linewidth]{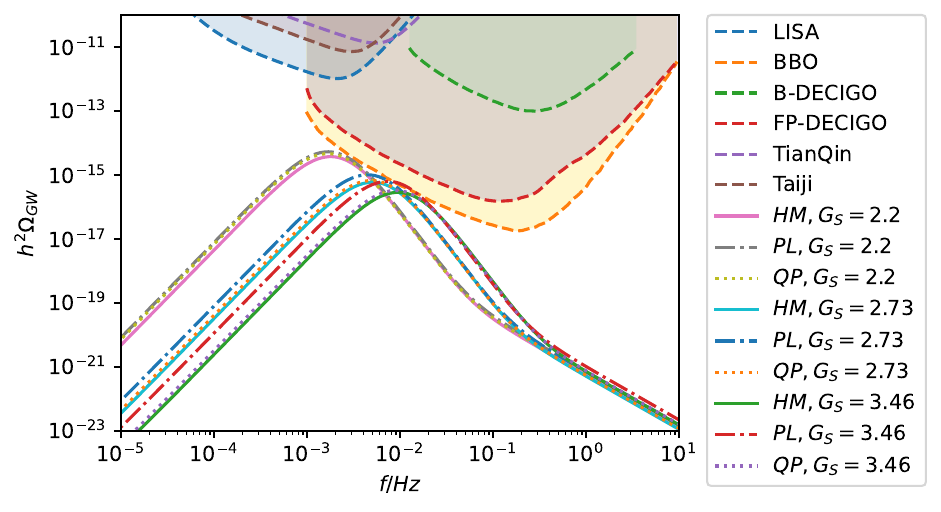}
    \includegraphics[width=0.49\linewidth]{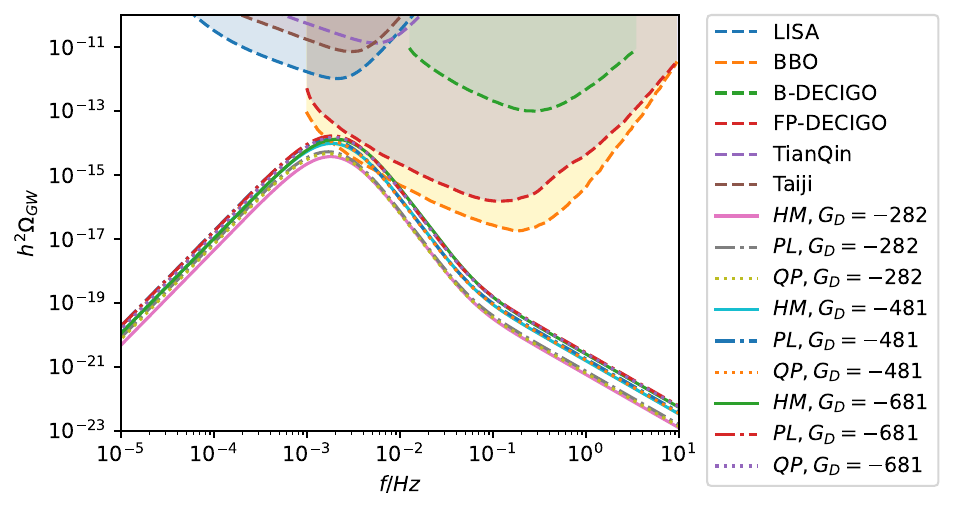}
    \includegraphics[width=0.49\linewidth]{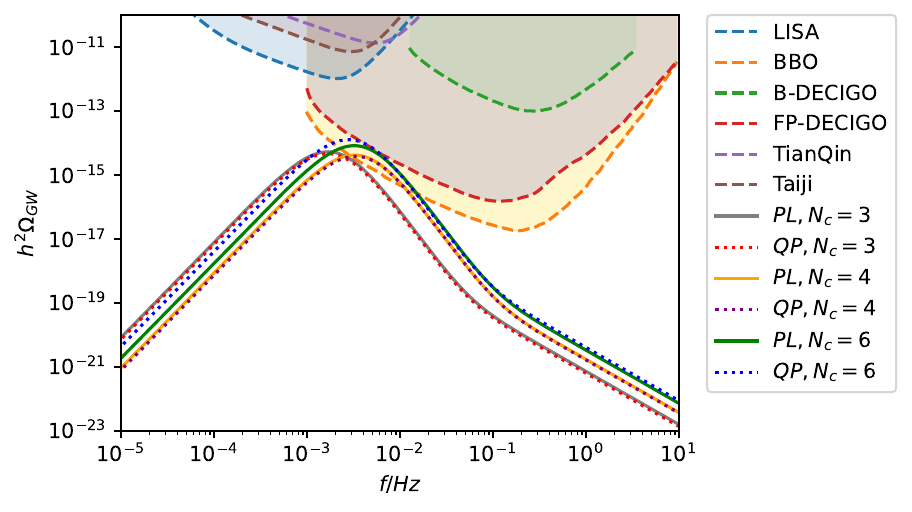}
    \includegraphics[width=0.49\linewidth]{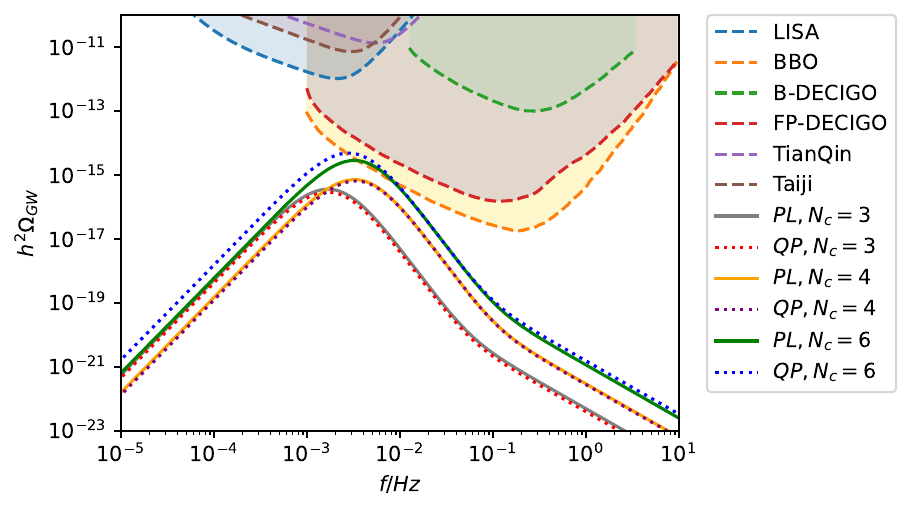}
    \caption{Gravitational wave spectra and detector sensitivity curves for different PL models. In the upper panels, $N_c=3$: the upper-left panel shows the results for fixed $G_D=-282$ with varying $G_S=2.2,\,2.73,\,3.46$, while the upper-right panel presents the results for fixed $G_S=2.2$ with varying $G_D=-282,\,-481,\,-681$. In the lower-left panel, $G_D=-282$ and $G_S=2.2$ are fixed, while $N_c=3,\,4,\,6$ are considered. The lower-right panel shows the gravitational-wave spectra for $N_c=3,\,4,\,6$ including the Standard Model contribution. The solid, dash-dotted, and dotted lines correspond to the Haar-measure potential, polynomial potential, and quasi-particle model, respectively.}
    \label{chiralgw4}
\end{figure}



The gravitational wave spectra from the chiral phase transition for different PL models and choices of model parameters are summarized in Fig.~\ref{chiralgw4}. In the upper-left panel, $N_c=3$ and $G_D=-282$ are fixed while $G_S$ is varied over $G_S=2.2,\,2.73,\,3.46$, leading to a change in the gravitational-wave amplitude of nearly two orders of magnitude. In the upper-right panel, $N_c=3$ and $G_S=2.2$ are fixed and $G_D$ is varied over $G_D=-282,\,-481,\,-681$, which results in an amplitude variation of about one order of magnitude. In the lower-left panel, the couplings are fixed at $G_S=2.2$ and $G_D=-282$, and the color number is varied as $N_c=3,\,4,\,6$, again giving rise to an amplitude change of roughly one order of magnitude. In the lower-right panel, the spectra for $N_c=3,\,4,\,6$ including the Standard Model contribution are shown, where the overall amplitude is suppressed by nearly two orders of magnitude compared with the dark-matter–dominated case.
These variations are significantly larger than those caused by including the PL kinetic term, confirming that the latter plays only a subleading role in the chiral phase transition.

\section{conclusion and discussion}\label{conclution}
The QCD phase transition in the early Universe involves both the confinement and chiral transitions. For the confinement transition, properly accounting for the kinetic term of the Polyakov loop (PL), which serves as the order parameter, rather than assuming a canonical normalization, is crucial for obtaining reliable observable signals. This has two important implications: the PL kinetic term requires a nontrivial, field-dependent renormalization factor instead of being treated as unity, and in phase transitions involving multiple order parameters, a consistent treatment demands that all corresponding kinetic terms be included.

We derive the renormalization factor of the PL kinetic term and, combined with three forms of the PL effective potential—the Haar-measure, polynomial, and quasiparticle models—compute the gravitational-wave spectra from the confinement transition. Treating the kinetic term as trivial or nontrivial can lead to order-of-magnitude differences in the predicted gravitational-wave amplitude. By contrast, for the chiral phase transition, we consider the full effective action, including the kinetic terms of both fermions and the PL, together with the PNJL effective potential, where the PL sector is again described by the same three models. In this case, whether or not the PL kinetic term is included has only a minor impact on the gravitational-wave spectrum, indicating that the transition dynamics is dominated by the fermionic order parameter. Consequently, the chiral transition can in practice be well approximated as a single-field problem.

However, a key question remains to be addressed. As noted in the section~\ref{Conf}, our discussion has so far been limited to the kinetic terms of the Polyakov loop derived from the classical pure Yang–Mills Lagrangian; the quantum corrections to these terms have not yet been taken into account. Several studies have examined quantum corrections to pure Yang–Mills kinetic terms using non‑perturbative methods, such as the effective model~\cite{Wirstam:2001ka, Dumitru:2002cf} and the FRG method~\cite{Guan:2025mce}, and modified kinetic term is Eq.\eqref{termkaa} with a non-trivial renormalization factor contributing from the quantum corrections
\begin{equation}
    K(A)\rightarrow Z_{Q}(A)K(A)=Z_Q[A(l)]Z_l\vec{\nabla}l\cdot\vec{\nabla}l.
\end{equation}
These corrections are generally non‑trivial and could significantly influence the predicted gravitational‑wave signals from confinement phase transitions. Consequently, future work should incorporate such quantum effects.

\noindent {\bf{Acknowledgements}}

\appendix
\section{The Proof of the Equal Eigenvalue for $SU(3)$ Yang-Mills Theory}\label{appA}

In this appendix, we will prove the equal eigenvalue assumption within $SU(3)$ pure Yang-Mills Theory. Let us start from the general form of the Polyakov loop under the Polyakov gauge Eq.\eqref{matex} which requires the Polyakov loop to be diagonal with
\begin{equation}\label{eq:PolGauge}
    \sum_{i=1}^{N_c} q_i=0,
\end{equation}
One can immediately noticed that the equal eigenvalue conditions is true for $N_c=2$. Under this gauge, the trace of the Polyakov loop is a complex field. However, in the three models we consider, even if we set the field $l$ as a complex field, the path that ultimately minimizes the effective potential will always satisfy $ l^* =l$. These properties come from the $Z_3$ center symmetry of the $SU(3)$ gauge group. As a consequence, if the trace of Polyakov loops is real, then the three eigenphases $q_i$ would satisfy another condition
\begin{equation}
    {\rm Im}(e^{iq_q}+e^{iq_2}+e^{iq_3})=\sin(q_1)+\sin(q_2)-\sin(q_1+q_2)=0,
\end{equation}
where we have used the condition Eq.\eqref{eq:PolGauge}. If we set $\sin q_2=x$, the above equation would become
\begin{equation}
    \sin q_1 + (1- \cos q_1)x=\sin q_1 \sqrt{1-x^2}.
\end{equation}
As long as $q_1\neq0$, this equation admit two solutions which is given by
\begin{equation}
    x_1=0\quad x_2=-\sin q_1,
\end{equation}
for the first solution $\sin q_2=x_1$, if we require $-\pi\leq q_2\leq\pi$, we have $q_2=0$~\footnote{Another solution $q_2=\pi$ is dropped because, if $q_2=\pi$, then $\cos q_2\neq\sqrt{1-\sin q_2}$.} and Eq.\eqref{eq:PolGauge} immediately require $q_1=-q_3=q$ which is exactly the Eq.\eqref{eq:eeconif} shown. If $\sin q_2=x_2=-\sin q_1$, then from  Eq.\eqref{eq:PolGauge} we obtained that $q_1=q$, $q_2=-q$ and $q_3=0$, which is just equivalent form of the the equal eigenvalue conditions Eq.\eqref{eq:eeconif}. As a conclusion, we proved the equal eigenvalue conditions for $N_c\leq3$ always satisfied.

\section{Derivation of renormalization factor of Fermion under the equal eigenvalue assumption}\label{appendixd}
The effective action including quantum corrections is the generating functional of $n$-point 1-PI correlation functions. By performing a Fourier transform on the fields and correlation functions and applying a derivative expansion / zero-momentum expansion, the terms quadratic in derivatives correspond to the kinetic terms. The coefficient of the kinetic term is given by
\begin{equation}
    Z_\sigma^{-1}=-\frac{d\Gamma_{\sigma\sigma}(q_0,\vec{q},\sigma)}{d\vec{q}^2}|_{q_0=0,\vec{q}^2=0},
\end{equation}
which receives contributions from quasiquarks. Since this work considers thermodynamic tunneling at finite temperature, the renormalization factor should also be computed within the three-dimensional framework. In this case, the calculation should also include the background field $A_4$, just as in the computation of the effective potential.

Since $\Gamma_{\sigma\sigma}$ is the Fourier transform of the 1-PI two-point correlation function, i.e., the Feynman diagrams with two external 1-PI legs, to compute it we need to include $A_4$ to the Lagrangian in Eq.~(\ref{sigmaNJL}),
\begin{equation}
    \mathcal{L}^{eff}_{NJL}=\bar{q}[i\slashed\partial-M(\sigma)+\gamma^0(\mu-iA_4)]q-V_{NJL}^{tree},
\end{equation}
and to collect the relevant couplings from the above Lagrangian. The Feynman rules involving the $\sigma$ field are then obtained as follows:\\
\tikzset{every picture/.style={line width=0.75pt}} 

\begin{tikzpicture}[x=0.75pt,y=0.75pt,yscale=-1,xscale=1]

\draw  [dash pattern={on 4.5pt off 4.5pt}]  (22.91,101.58) -- (96.6,101.58) ;
\draw  [dash pattern={on 4.5pt off 4.5pt}]  (26.02,171.58) -- (66.24,171.58) ;
\draw    (66.24,171.58) -- (103.46,205.62) ;
\draw    (66.24,171.58) -- (102.72,144.35) ;
\draw    (241.18,156.83) -- (314.87,192.38) ;
\draw  [dash pattern={on 4.5pt off 4.5pt}]  (241.18,193.14) -- (314.87,157.59) ;

\draw (10.21,89.95) node [anchor=north west][inner sep=0.75pt]    {$\sigma $};
\draw (95.82,89.95) node [anchor=north west][inner sep=0.75pt]    {$\sigma $};
\draw (16.32,160.71) node [anchor=north west][inner sep=0.75pt]    {$\sigma$};
\draw (229.97,183.03) node [anchor=north west][inner sep=0.75pt]    {$\sigma $};
\draw (314.83,145.96) node [anchor=north west][inner sep=0.75pt]    {$\sigma $};
\draw (101.67,195) node [anchor=north west][inner sep=0.75pt]    {$\psi $};
\draw (101.67,132.96) node [anchor=north west][inner sep=0.75pt]    {$\psi $};
\draw (315.32,184.03) node [anchor=north west][inner sep=0.75pt]    {$\psi $};
\draw (228.22,148.47) node [anchor=north west][inner sep=0.75pt]    {$\psi $};
\draw (110.45,88.79) node [anchor=north west][inner sep=0.75pt]    {$=-\frac{3i}{4G_{S}} -\frac{3iG_{D}}{64G_{S}^{3}}\sigma$};
\draw (116.04,161.57) node [anchor=north west][inner sep=0.75pt]    {$=i+\frac{iG_{D}}{32G_{S}^{2}}\sigma$};
\draw (336.11,163.73) node [anchor=north west][inner sep=0.75pt]    {$=\frac{iG_{D}}{32G_{S}^{2}}$};
\end{tikzpicture}\\
The diagram in the first row represents the tree-level two-point correlation function. Since it contain no $q^2$ terms, the tree-level kinetic term is zero.  The second row shows the Feynman diagrams for the interaction between the $\sigma$ field and quarks, which contribute to the loop diagrams shown below. In the mean-field approximation, the self-interactions of the $\sigma$ field do not need to be considered. Thus, the one-loop two-point correlation function has the following three contributions:
\tikzset{every picture/.style={line width=0.75pt}} 

\begin{tikzpicture}[x=0.75pt,y=0.75pt,yscale=-1,xscale=1]

\draw  [dash pattern={on 4.5pt off 4.5pt}]  (49,150) -- (199,150) ;
\draw   (302,150) .. controls (302,136.19) and (313.19,125) .. (327,125) .. controls (340.81,125) and (352,136.19) .. (352,150) .. controls (352,163.81) and (340.81,175) .. (327,175) .. controls (313.19,175) and (302,163.81) .. (302,150) -- cycle ;
\draw  [dash pattern={on 4.5pt off 4.5pt}]  (352,150) -- (399,150) ;
\draw  [dash pattern={on 4.5pt off 4.5pt}]  (252,150) -- (302,150) ;
\draw  [dash pattern={on 4.5pt off 4.5pt}]  (451,173) -- (601,173) ;
\draw   (501,148) .. controls (501,134.19) and (512.19,123) .. (526,123) .. controls (539.81,123) and (551,134.19) .. (551,148) .. controls (551,161.81) and (539.81,173) .. (526,173) .. controls (512.19,173) and (501,161.81) .. (501,148) -- cycle ;

\draw (50,128.4) node [anchor=north west][inner sep=0.75pt]    {$\sigma $};
\draw (192,129.4) node [anchor=north west][inner sep=0.75pt]    {$\sigma $};
\draw (252,128.4) node [anchor=north west][inner sep=0.75pt]    {$\sigma $};
\draw (387,129.4) node [anchor=north west][inner sep=0.75pt]    {$\sigma $};
\draw (451,150.4) node [anchor=north west][inner sep=0.75pt]    {$\sigma $};
\draw (588,150.4) node [anchor=north west][inner sep=0.75pt]    {$\sigma $};
\draw (320,103.4) node [anchor=north west][inner sep=0.75pt]    {$\psi $};
\draw (320,153.4) node [anchor=north west][inner sep=0.75pt]    {$\psi $};
\draw (518,100.4) node [anchor=north west][inner sep=0.75pt]    {$\psi $};
\end{tikzpicture}.\\
Direct calculation yields the following result:
\begin{equation}\label{1PI}
    \Gamma_{\sigma\sigma}(q)=-\frac{3}{4G_S}-\frac{3G_D}{64G_S^3}\sigma+3N_c(1+\frac{G_D\sigma}{32G_S^2})^2A(q^2)+3N_c\frac{G_D}{32G_S^2}B(q^2),
\end{equation}
here, $A(q^2)$ and $B(q^2)$ are the one-loop functions corresponding to the second and third diagrams, respectively, and are given by
\begin{equation}
\begin{split}
    &A(q^2)=\frac{1}{N_c}\Tr_c\int\frac{d^4k}{i(2\pi)^4}\frac{\Tr[(\slashed k+\slashed q+M(\sigma))(\slashed k+M(\sigma))]}{[(k+q)^2-M^2(\sigma)][k^2-M^2(\sigma)]},
    \\
    &B(q^2)=\frac{1}{N_c}\Tr_c\int\frac{d^4k}{i(2\pi)^4}\frac{M(\sigma)}{k^2-M^2(\sigma)}.
\end{split}
\end{equation}
Since the $\sigma$ self-energy corrections considered in this paper arise from both quantum and thermal fluctuations, the renormalization factor must be calculated using finite-temperature field theory, with $k = (k_0 + \mu - i A_4, \vec{k})$. Consequently, the four-momentum integral should be replaced by the following sum:
\begin{equation}
    \int\frac{d^4k}{(2\pi)^4}\rightarrow\int_T\frac{d^4k}{(2\pi)^4}\equiv iT\sum_n\int\frac{d^3\vec{k}}{(2\pi)^3}.
\end{equation}
It can be seen that the bubble diagrams do not depend on the external momentum $\mathbf{q}$, so the only nonzero contribution to the renormalization factor~(\ref{1PI}) comes from the one-loop function $A(q^2)$.

Using the following technique, the part of $A(q^2)$ that depends on the external momentum can be further isolated. For this purpose, we first write the loop function as
\begin{equation}
\begin{split}
    A(q^2)=&\frac{1}{N_c}\Tr_c\int_T\frac{d^4k}{i(2\pi)^4}\frac{4[k^2+k\cdot q+M^2(\sigma)]}{[(k+q)^2-M^2(\sigma)][k^2-M^2(\sigma)]}
    \\
    =&\frac{1}{N_c}\Tr_c\int_T\frac{d^4k}{i(2\pi)^4}\frac{4[(k+q)^2-M^2(\sigma)]-2[q^2-4M^2(\sigma)]-2[q^2+2k\cdot q]}{[(k+q)^2-M^2(\sigma)][k^2-M^2(\sigma)]}.
\end{split}
\end{equation}
The last term in the numerator can be written as $2[(k+q)^2 - M^2(\sigma)] - 2[k^2 - M^2(\sigma)]$. By changing the internal variable as $d^4k \rightarrow d^4(k - q)$, the integrals of these two parts cancel each other. Thus, the final result of the loop function is
\begin{equation}\label{loopA}
    A(q^2)=\frac{4}{N_c}\Tr_c\int_T\frac{d^4k}{i(2\pi)^4}\frac{1}{k^2-M^2(\sigma)}-2[q^2-4M^2(\sigma)]I(q,\sigma),
\end{equation}
where $I(q, \sigma)$ is the generating function of the loop integral,
\begin{equation}
    I(q,\sigma)=\frac{1}{N_c}\Tr_c\int_T\frac{d^4k}{i(2\pi)^4}\frac{1}{[(k+q)^2-M^2(\sigma)][k^2-M^2(\sigma)]}.
\end{equation}
The first term in Eq.~\eqref{loopA} is independent of the external momentum and does not contribute to the renormalization factor. Therefore, only the second term needs to be considered, yielding the final expression for the renormalization factor,
\begin{equation}
    Z_\sigma^{-1}=2N_f N_c (1+\frac{G_D\sigma}{32G_S^2})^2[I(0,\sigma)+4M^2(\sigma)\frac{dI(0,\sigma)}{d\vec{q}^2}].
\end{equation}
In the following, we outline the evaluation of the finite-temperature loop function $I(q,\sigma)$.

To include the effect of the temporal background $A_0 = i A_4$, we define the generalized chemical potential $\hat{\mu} = \mu + i A_4$ and replace $k_0 = i(\omega_n - i \hat{\mu})$ with $\omega_n = (2n+1)\pi T$. The integral then becomes
\begin{equation}
\begin{split}
    I(q,\sigma)=\frac{1}{N_c}\Tr_c\int\frac{d^3\vec{k}}{(2\pi)^3}\sum_{n=-\infty}^\infty\frac{1}{[(2n+1)\pi T-i\hat{\mu}-iq_0]^2+E^2_{k+q}}
    \\
    \times \frac{1}{([(2n+1)\pi T-i\hat{\mu}]^2+E^2_k)},
\end{split}
\end{equation}
where $E_k^2 = \vec{k}^2 + M^2(\sigma)$. Since the renormalization factor is calculated in the limit $q_0 \rightarrow 0$, this limit can be applied before performing the sum over $n$, yielding
\begin{equation}
\begin{split}
    I(0,\vec{q},\sigma)=&\frac{T}{N_c}\Tr_c\int\frac{d^3\vec{k}}{(2\pi)^3}
    \\
    \times&\frac{E_k[{\rm Tanh}(\frac{\hat{\mu}-E_{k+q}}{T})-{\rm Tanh}(\frac{\hat{\mu}+E_{k+q}}{T})]-E_{k+q}[{\rm Tanh}(\frac{\hat{\mu}-E_{k}}{T})-{\rm Tanh}(\frac{\hat{\mu}+E_{k}}{T})]}{4TE_kE_{k+q}(E_{k+q}+E_k)(E_{k+q}-E_k)}.
\end{split}
\end{equation}
One can conveniently define the effective distribution functions $F_+$ and $F_-$:
\begin{equation}
    F_+(\omega)=\frac{2e^{-\frac{\omega-\hat{\mu}}{T}}}{1+e^{-\frac{\omega-\hat{\mu}}{T}}},\quad\quad\quad\quad F_-(\omega)=\frac{2e^{-\frac{\omega+\hat{\mu}}{T}}}{1+e^{-\frac{\omega+\hat{\mu}}{T}}}.
\end{equation}
The loop integral can then be expressed using these distribution functions:
\begin{equation}
\begin{split}
    I(0,\vec{q},\sigma)=\frac{\Tr_c}{N_c}\int\frac{d^3\vec{k}}{(2\pi)^3}&\frac{1}{4E_kE_{k+q}}\Bigg[\frac{F_+(E_{k+q})-F_+(E_k)+F_-(E_{k+q})-F_-(E_k)}{E_{k+q}-E_k}
    \\
    &+\frac{2-F_+(E_{k+q})-F_-(E_{k+q})-F_+(E_k)-F_-(E_k)}{E_{k+q}+E_k}\Bigg],
\end{split}
\end{equation}
the trace over color indices acts only on the effective distribution functions.

In $SU(3)$ gauge theory, the background gauge field in the Polyakov gauge is written as a diagonal matrix~\eqref{matex}. Therefore, the trace over the effective distribution functions can be taken separately to obtain the two Polyakov loop functions,
\begin{equation}
\begin{split}
   f_+(\omega,r)&=\frac{\Tr_c}{N_c}F_+(\omega,r)
   \\
   &=\frac{e^{-\frac{\Omega_+}{T}}}{N_c}\Tr_c\bigg[\diag\bigg(e^{i2\pi q_1},e^{i2\pi q_2},...,e^{i2\pi q_{N_c}}\bigg)
   \\
   &\indent\times\diag\bigg(\frac{1}{1+e^{-\frac{\Omega_+}{T}}e^{i2\pi q_1}},\frac{1}{1+e^{-\frac{\Omega_+}{T}}e^{i2\pi q_2}},...,\frac{1}{1+e^{-\frac{\Omega_+}{T}}e^{i2\pi q_{N_c}}}\bigg)\bigg]
   \\
   &=\frac{1}{N_c}\sum_{j=1}^{N_c}\frac{1}{1+e^{\frac{\Omega_+}{T}}e^{-i2\pi q_j}},
\end{split}
\end{equation}
and 
\begin{equation}
\begin{split}
    f_-(\omega,r)&=\frac{\Tr_c}{N_c}F_-(\omega,r)
   \\
   &=\frac{e^{-\frac{\Omega_-}{T}}}{N_c}\Tr_c\bigg[\diag\bigg(e^{-i2\pi q_1},e^{-i2\pi q_2},...,e^{-i2\pi q_{N_c}}\bigg)
   \\
   &\indent\times\diag\bigg(\frac{1}{1+e^{-\frac{\Omega_-}{T}}e^{-i2\pi q_1}},\frac{1}{1+e^{-\frac{\Omega_-}{T}}e^{-i2\pi q_2}},...,\frac{1}{1+e^{-\frac{\Omega_-}{T}}e^{-i2\pi q_{N_c}}}\bigg)\bigg]
   \\
   &=\frac{1}{N_c}\sum_{j=1}^{N_c}\frac{1}{1+e^{\frac{\Omega_-}{T}}e^{i2\pi q_j}},
\end{split}
\end{equation}
where $\Omega_\pm \equiv (\omega \pm \mu)/T$, and $q_j$ is expressed in terms of $r$ under the equal eigenvalue conditions. If the effect of the chemical potential is neglected, then $\Omega_+ = \Omega_- = \omega$. Furthermore, since each nonzero $q_i$ has a corresponding $q_j = -q_i$, $f_+$ and $f_-$ become identical,
\begin{equation}
    f_+(\omega,r)=f_-(\omega,r)=\frac{1}{N_c}\sum_{j=1}^{N_c}\frac{1}{1+e^{\frac{\omega}{T}}e^{i2\pi q_j}}=f(\omega,r).
\end{equation}
Finally, the loop function can be expressed as a complex function of $\sigma$ and $r$:
\begin{equation}
\begin{split}
    I(0,\vec{q},\sigma)=\int\frac{d^3\vec{k}}{(2\pi)^3}\frac{1}{2E_kE_{k+q}}\Bigg[\frac{f(E_{k+q},r)-f(E_k,r)}{E_{k+q}-E_k}+\frac{1-f(E_{k+q},r)-f(E_k,r)}{E_{k+q}+E_k}\Bigg].
\end{split}
\end{equation}
From this expression, $I(0, \vec{q} = 0, \sigma)$ can be derived:
\begin{equation}
    \lim_{q\to 0}\frac{f(E_{k+q},r)-f(E_k,r)}{E_{k+q}-E_k}=\frac{\partial f(E_k,r)}{\partial E_k},
\end{equation}
thus
\begin{equation}
    I(0,\sigma)=\lim_{q\to 0}I(0,\vec{q},\sigma)=\int\frac{d^3\vec{k}}{(2\pi)^3}\frac{1}{2E_k^2}\Bigg[\frac{\partial f(E_k,r)}{\partial E_k}+\frac{1-2f(E_k,r)}{2E_k}\Bigg].
\end{equation}
Next, we derive $\frac{dI(0, \vec{q}, \sigma)}{d\vec{q}^2}\Big|_{\vec{q}=0}$. One can define
\begin{equation}
\begin{split}
    a(q^{2})&=\frac{1}{2E_{k}E_{k+q}}\frac{1}{E_{k+q}-E_{k}}\left[f(E_{k+q})-f(E_{k})\right],
    \\
    b(q^{2})&=\frac{1}{2E_{k}E_{k+q}}\frac{1}{E_{k+q}+E_{k}}\left[1-f(E_{k+q})-f(E_{k})\right].
\end{split}
\end{equation}
Then, $I(0, \vec{q}, \sigma)$ can be expressed as
\begin{equation}
    I(q^2)=\int\frac{d^3\vec k}{(2\pi)^3}[a(q^2)+b(q^2)]
\end{equation}
The derivative with respect to $q^2$ can be interchanged with the integral over $\vec{k}$, so we first differentiate $a(q^2)$ and $b(q^2)$ separately,
\begin{equation}
    \begin{split}
        a'(q^{2})=\bigg[-\frac{\left(E_{k+q}+E_{k+q}-E_{k}\right)}{2E_{k}\left[E_{k+q}\left(E_{k+q}-E_{k}\right)\right]^{2}}&\left(f(E_{k+q})-f(E_{k})\right)
        \\
        +\frac{1}{2E_{k}E_{k+q}}&\frac{1}{E_{k+q}-E_{k}}\frac{\partial f(E_{k+q},r)}{\partial E_{k+q}}\bigg]\frac{dE_{k+q}}{d(q^{2})}
        \\
        b'(q^{2})=-\bigg[\frac{\left(E_{k+q}+E_{k+q}+E_{k}\right)}{2E_{k}\left[E_{k+q}\left(E_{k+q}+E_{k}\right)\right]^{2}}&\left(1-f(E_{k+q})-f(E_{k})\right)
        \\
        +\frac{1}{2E_{k}E_{k+q}}&\frac{1}{E_{k+q}+E_{k}}\left(-\frac{\partial f(E_{k+q},r)}{\partial E_{k+q}}\right)\bigg]\frac{dE_{k+q}}{d(q^{2})},
    \end{split}
\end{equation}
and let $q\to 0$,
\begin{equation}
\begin{split}
    a'(0)=&\lim_{q\to 0}a'(q^2)=0
    \\
    b'(0)=&\lim_{q\to 0}b'(q^2)=\lim_{q\to0}\frac{dE_{k+q}}{d(q^2)}\frac{1}{8E_k^4}\bigg[-3(1-2f(E_k))-2E_k\frac{\partial f(E_{k+q},r)}{\partial E_{k+q}}\bigg],
\end{split}
\end{equation}
then
\begin{equation}
    I'(0)=\lim_{q\rightarrow0}\int\frac{d^{3}\vec k}{(2\pi)^{3}}\frac{dE_{k+q}}{d(q^{2})}\frac{1}{8E_{k}^{4}}\left[-3\left(1-2f(E_{k})\right)-2E_{k}\frac{\partial f(E_{k+q},r)}{\partial E_{k+q}}\right].
\end{equation}
Note that the subscript $k+q$ involves a vector sum, thus 
\begin{equation}
    \frac{dE_{k+q}}{d(q^{2})}=\frac{1}{2}\frac{q+k\cos\theta_{\vec{k},\vec{q}}}{qE_{k+q}}.
\end{equation} 
after integrating over $d^{3}\vec k$, the angular-dependent part vanishes, so finally
\begin{equation}
    I'(0)=\int\frac{d^3\vec{k}}{(2\pi)^{3}}\frac{1}{16E_{k}^{5}}\left[6f(\omega)-2E_{k}\frac{\partial f(\omega,r)}{\partial \omega}-1\right].
\end{equation}
In the calculation of $dI/d\vec{q^2}$, the result from \cite{Reichert:2021cvs} is used: Lorentz symmetry in the zero-temperature theory requires replacing $dI/d\vec{q^2}$ with $dI/d\vec{q^2} + \int \frac{d^3\vec{k}}{(2\pi)^3} \frac{1}{8 E_k^5}$ to ensure the positivity of the renormalization factor. After performing the numerical integration over momentum space, the wavefunction factor $Z(\sigma, r, T)$ is obtained.

\section{Bounce equation for multi non-trivial kinetic term}\label{appc}
In this section, we derive the general multi-field bounce equation, which is used in Sec.~\ref{secivd}. The D-dimension Euclidean action that contains multi non-trivial kinetic term reads:
\begin{equation}
    S=\frac{2\pi^{\frac{D}{2}}}{\Gamma(\frac{D}{2})}\int_0^\infty dr\,r^{D-1}\left[\frac{Z_i^{-1}}{2}\left(\frac{d\phi_i}{dr}\right)^2+V_{eff}(\overrightarrow{\phi} )\right],
\end{equation}
here, the repeated indices $i$ imply summation. To obtain the bounce equation, one can apply the Euler–Lagrange equation:
\begin{equation}
    \frac{d}{dr}\left[r^{D-1}Z_i^{-1}\frac{d\phi_i}{dr}\right]-\frac{\partial}{\partial\phi_i}\left[r^{D-1}V_{eff}(\overrightarrow{\phi} )+r^{D-1}\sum_j\frac{Z_j^{-1}}{2}\left(\frac{d\phi_j}{dr}\right)^2\right]=0,
\end{equation}
here and in this section, repeated indices DO NOT imply summation; only $\sum$ denotes summation, unless otherwise specified. One can get the $i$th equation
\begin{equation}\label{itheq}
    \frac{d^2\phi_i}{dr^2}+\frac{D-1}{r}\frac{d\phi_i}{dr}-\frac{1}{Z_i}\frac{dZ_i}{dr}\frac{d\phi_i}{dr}+\frac{1}{2}\sum_j\frac{Z_i}{Z_j^2}\frac{\partial Z_j}{\partial \phi_i}\left(\frac{d\phi_j}{dr}\right)^2-Z_i\frac{\partial V_{eff}}{\partial \phi_i}=0.
\end{equation}
To solve the multi-field bounce equations, we need to determine the correct tunneling path, for which we adopt the path deformation method. This requires decomposing the equations into components tangential and normal to the path. Assuming an arbitrary path parametrized as $\overrightarrow{\phi}(x)$, substituting it into the equations of motion yields a vector form:
\begin{equation}\label{totaleom}
\begin{split}
    \frac{d^2x}{dr^2}\frac{d\overrightarrow{\phi}}{dx}+\frac{d^2\overrightarrow{\phi}}{dx^2}&\left(\frac{dx}{dr}\right)^2+\frac{D-1}{r}\frac{d\overrightarrow{\phi}}{dx}\frac{dx}{dr} -\sum_i \frac{1}{Z_i}\frac{dZ_i}{dx}\frac{d\phi_i}{dx}\left(\frac{dx}{dr}\right)^2\overrightarrow{e}_i 
    \\
    &+\frac{1}{2}\sum_i \sum_j \frac{\partial Z_j}{\partial \phi_i}\frac{Z_i}{Z_j^2}\left(\frac{d\phi_j}{dx}\right)^2\left(\frac{dx}{dr}\right)^2 \overrightarrow{e}_i -\sum_i Z_i \frac{\partial V_{eff}}{\partial \phi_i}\overrightarrow{e}_i= 0.
\end{split}
\end{equation}
For convenience in calculation, we set $\left|\frac{d\overrightarrow{\phi}}{dx}\right|=1$, and then one can find $\frac{d^2\overrightarrow{\phi}}{dx^2}\cdot\frac{d\overrightarrow{\phi}}{dx}=0$. Therefore, the equation tangential to the path can be obtained via \eqref{totaleom}$\cdot\frac{d\overrightarrow{\phi}}{dx}$:
\begin{equation}\label{paralleleq}
    \begin{split}
        \frac{d^2x}{dr^2}+\frac{D-1}{r}\frac{dx}{dr}&-\sum_i \frac{1}{Z_i}\frac{dZ_i}{dx}\left(\frac{d\phi_i}{dx}\right)^2\left(\frac{dx}{dr}\right)^2
        \\
        &+\frac{1}{2}\sum_i\sum_j\frac{Z_i}{Z_j^2}\frac{\partial Z_j}{\partial \phi_i}\frac{d\phi_i}{dx}\left(\frac{d\phi_j}{dx}\right)^2\left(\frac{dx}{dr}\right)^2 - \sum_i Z_i \frac{d\phi_i}{dx}\frac{\partial V_{eff}}{\partial \phi_i}=0.
    \end{split}
\end{equation}
The barrier here is found by solving $\sum_i Z_i \frac{d\phi_i}{dx}\frac{\partial V_{eff}}{\partial \phi_i}=0$. The normal part of Eq.\eqref{totaleom} can be obtained via \eqref{totaleom}$-$\eqref{paralleleq}$\frac{d\overrightarrow{\phi}}{dx}$:
\begin{equation}
\begin{split}
    \bigg[\frac{d^2\overrightarrow{\phi}}{dx^2} -\sum_i \frac{1}{Z_i}\frac{dZ_i}{dx}\frac{d\phi_i}{dx}\overrightarrow{e}_i +&\sum_i\sum_j \frac{1}{Z_j}\frac{dZ_j}{dx}\left(\frac{d\phi_j}{dx}\right)^2 \frac{d\phi_i}{dx}\overrightarrow{e}_i
    \\
    +\frac{1}{2}\sum_i \sum_j \frac{\partial Z_j}{\partial \phi_i}\frac{Z_i}{Z_j^2}\left(\frac{d\phi_j}{dx}\right)^2 \overrightarrow{e}_i -&\frac{1}{2}\sum_i\sum_j\sum_k\frac{Z_k}{Z_j^2}\frac{\partial Z_j}{\partial \phi_k}\frac{d\phi_k}{dx}\left(\frac{d\phi_j}{dx}\right)^2\frac{d\phi_i}{dx}\overrightarrow{e}_i\bigg]\left(\frac{dx}{dr}\right)^2
    \\
    -&\sum_i Z_i \frac{\partial V_{eff}}{\partial \phi_i}\overrightarrow{e}_i +\sum_i\sum_j Z_j \frac{d\phi_j}{dx}\frac{\partial V_{eff}}{\partial \phi_j}\frac{d\phi_i}{dx}\overrightarrow{e}_i =0.
\end{split}
\end{equation}
The left–hand side of the equation is identified as the normal force $N$. When the path is deformed such that $N \to 0$, the correct tunneling path is obtained. The method described in this section has been implemented in our numerical package \texttt{VacuumTunneling}~\cite{Hua:2025fap}, which can be used to solve the general multi-field bounce equations.

\bibliographystyle{unsrt}  
\bibliography{main} 

\end{document}